\documentclass[]{aastex631}
\usepackage{rotating}
\usepackage{xspace}
\usepackage{booktabs}
\usepackage{amsmath}
\usepackage{longtable}
\usepackage{booktabs} 
\usepackage{multirow}

\newcommand{\sigmas}{\ensuremath{\sigma}\xspace}
\shorttitle{Evidence for Neutrino Emission from X-ray Bright Active Galactic Nuclei with IceCube}
\shortauthors{IceCube Collaboration}

\begin{document}

\title{Evidence for Neutrino Emission from X-ray Bright Active Galactic Nuclei with IceCube}


\affiliation{III. Physikalisches Institut, RWTH Aachen University, D-52056 Aachen, Germany}
\affiliation{Department of Physics, University of Adelaide, Adelaide, 5005, Australia}
\affiliation{Dept. of Physics and Astronomy, University of Alaska Anchorage, 3211 Providence Dr., Anchorage, AK 99508, USA}
\affiliation{School of Physics and Center for Relativistic Astrophysics, Georgia Institute of Technology, Atlanta, GA 30332, USA}
\affiliation{Dept. of Physics, Southern University, Baton Rouge, LA 70813, USA}
\affiliation{Dept. of Physics, University of California, Berkeley, CA 94720, USA}
\affiliation{Lawrence Berkeley National Laboratory, Berkeley, CA 94720, USA}
\affiliation{Institut f{\"u}r Physik, Humboldt-Universit{\"a}t zu Berlin, D-12489 Berlin, Germany}
\affiliation{Fakult{\"a}t f{\"u}r Physik {\&} Astronomie, Ruhr-Universit{\"a}t Bochum, D-44780 Bochum, Germany}
\affiliation{Universit{\'e} Libre de Bruxelles, Science Faculty CP230, B-1050 Brussels, Belgium}
\affiliation{Vrije Universiteit Brussel (VUB), Dienst ELEM, B-1050 Brussels, Belgium}
\affiliation{Dept. of Physics, Simon Fraser University, Burnaby, BC V5A 1S6, Canada}
\affiliation{Department of Physics and Laboratory for Particle Physics and Cosmology, Harvard University, Cambridge, MA 02138, USA}
\affiliation{Dept. of Physics, Massachusetts Institute of Technology, Cambridge, MA 02139, USA}
\affiliation{Dept. of Physics and The International Center for Hadron Astrophysics, Chiba University, Chiba 263-8522, Japan}
\affiliation{Department of Physics, Loyola University Chicago, Chicago, IL 60660, USA}
\affiliation{Dept. of Physics and Astronomy, University of Canterbury, Private Bag 4800, Christchurch, New Zealand}
\affiliation{Dept. of Physics, University of Maryland, College Park, MD 20742, USA}
\affiliation{Dept. of Astronomy, Ohio State University, Columbus, OH 43210, USA}
\affiliation{Dept. of Physics and Center for Cosmology and Astro-Particle Physics, Ohio State University, Columbus, OH 43210, USA}
\affiliation{Niels Bohr Institute, University of Copenhagen, DK-2100 Copenhagen, Denmark}
\affiliation{Dept. of Physics, TU Dortmund University, D-44221 Dortmund, Germany}
\affiliation{Dept. of Physics and Astronomy, Michigan State University, East Lansing, MI 48824, USA}
\affiliation{Dept. of Physics, University of Alberta, Edmonton, Alberta, T6G 2E1, Canada}
\affiliation{Erlangen Centre for Astroparticle Physics, Friedrich-Alexander-Universit{\"a}t Erlangen-N{\"u}rnberg, D-91058 Erlangen, Germany}
\affiliation{Physik-department, Technische Universit{\"a}t M{\"u}nchen, D-85748 Garching, Germany}
\affiliation{D{\'e}partement de physique nucl{\'e}aire et corpusculaire, Universit{\'e} de Gen{\`e}ve, CH-1211 Gen{\`e}ve, Switzerland}
\affiliation{Dept. of Physics and Astronomy, University of Gent, B-9000 Gent, Belgium}
\affiliation{Dept. of Physics and Astronomy, University of California, Irvine, CA 92697, USA}
\affiliation{Karlsruhe Institute of Technology, Institute for Astroparticle Physics, D-76021 Karlsruhe, Germany}
\affiliation{Karlsruhe Institute of Technology, Institute of Experimental Particle Physics, D-76021 Karlsruhe, Germany}
\affiliation{Dept. of Physics, Engineering Physics, and Astronomy, Queen's University, Kingston, ON K7L 3N6, Canada}
\affiliation{Department of Physics {\&} Astronomy, University of Nevada, Las Vegas, NV 89154, USA}
\affiliation{Nevada Center for Astrophysics, University of Nevada, Las Vegas, NV 89154, USA}
\affiliation{Dept. of Physics and Astronomy, University of Kansas, Lawrence, KS 66045, USA}
\affiliation{Centre for Cosmology, Particle Physics and Phenomenology - CP3, Universit{\'e} catholique de Louvain, Louvain-la-Neuve, Belgium}
\affiliation{Department of Physics, Mercer University, Macon, GA 31207-0001, USA}
\affiliation{Dept. of Astronomy, University of Wisconsin{\textemdash}Madison, Madison, WI 53706, USA}
\affiliation{Dept. of Physics and Wisconsin IceCube Particle Astrophysics Center, University of Wisconsin{\textemdash}Madison, Madison, WI 53706, USA}
\affiliation{Institute of Physics, University of Mainz, Staudinger Weg 7, D-55099 Mainz, Germany}
\affiliation{Department of Physics, Marquette University, Milwaukee, WI 53201, USA}
\affiliation{Institut f{\"u}r Kernphysik, Universit{\"a}t M{\"u}nster, D-48149 M{\"u}nster, Germany}
\affiliation{Bartol Research Institute and Dept. of Physics and Astronomy, University of Delaware, Newark, DE 19716, USA}
\affiliation{Dept. of Physics, Yale University, New Haven, CT 06520, USA}
\affiliation{Columbia Astrophysics and Nevis Laboratories, Columbia University, New York, NY 10027, USA}
\affiliation{Dept. of Physics, University of Oxford, Parks Road, Oxford OX1 3PU, United Kingdom}
\affiliation{Dipartimento di Fisica e Astronomia Galileo Galilei, Universit{\`a} Degli Studi di Padova, I-35122 Padova PD, Italy}
\affiliation{Dept. of Physics, Drexel University, 3141 Chestnut Street, Philadelphia, PA 19104, USA}
\affiliation{Physics Department, South Dakota School of Mines and Technology, Rapid City, SD 57701, USA}
\affiliation{Dept. of Physics, University of Wisconsin, River Falls, WI 54022, USA}
\affiliation{Dept. of Physics and Astronomy, University of Rochester, Rochester, NY 14627, USA}
\affiliation{Department of Physics and Astronomy, University of Utah, Salt Lake City, UT 84112, USA}
\affiliation{Dept. of Physics, Chung-Ang University, Seoul 06974, Republic of Korea}
\affiliation{Oskar Klein Centre and Dept. of Physics, Stockholm University, SE-10691 Stockholm, Sweden}
\affiliation{Dept. of Physics and Astronomy, Stony Brook University, Stony Brook, NY 11794-3800, USA}
\affiliation{Dept. of Physics, Sungkyunkwan University, Suwon 16419, Republic of Korea}
\affiliation{Institute of Physics, Academia Sinica, Taipei, 11529, Taiwan}
\affiliation{Dept. of Physics and Astronomy, University of Alabama, Tuscaloosa, AL 35487, USA}
\affiliation{Dept. of Astronomy and Astrophysics, Pennsylvania State University, University Park, PA 16802, USA}
\affiliation{Dept. of Physics, Pennsylvania State University, University Park, PA 16802, USA}
\affiliation{Dept. of Physics and Astronomy, Uppsala University, Box 516, SE-75120 Uppsala, Sweden}
\affiliation{Dept. of Physics, University of Wuppertal, D-42119 Wuppertal, Germany}
\affiliation{Deutsches Elektronen-Synchrotron DESY, Platanenallee 6, D-15738 Zeuthen, Germany}

\author[0000-0001-6141-4205]{R. Abbasi}
\affiliation{Department of Physics, Loyola University Chicago, Chicago, IL 60660, USA}

\author[0000-0001-8952-588X]{M. Ackermann}
\affiliation{Deutsches Elektronen-Synchrotron DESY, Platanenallee 6, D-15738 Zeuthen, Germany}

\author{J. Adams}
\affiliation{Dept. of Physics and Astronomy, University of Canterbury, Private Bag 4800, Christchurch, New Zealand}

\author[0000-0002-9714-8866]{S. K. Agarwalla}
\altaffiliation{also at Institute of Physics, Sachivalaya Marg, Sainik School Post, Bhubaneswar 751005, India}
\affiliation{Dept. of Physics and Wisconsin IceCube Particle Astrophysics Center, University of Wisconsin{\textemdash}Madison, Madison, WI 53706, USA}

\author[0000-0003-2252-9514]{J. A. Aguilar}
\affiliation{Universit{\'e} Libre de Bruxelles, Science Faculty CP230, B-1050 Brussels, Belgium}

\author[0000-0003-0709-5631]{M. Ahlers}
\affiliation{Niels Bohr Institute, University of Copenhagen, DK-2100 Copenhagen, Denmark}

\author[0000-0002-9534-9189]{J.M. Alameddine}
\affiliation{Dept. of Physics, TU Dortmund University, D-44221 Dortmund, Germany}

\author[0009-0001-2444-4162]{S. Ali}
\affiliation{Dept. of Physics and Astronomy, University of Kansas, Lawrence, KS 66045, USA}

\author{N. M. Amin}
\affiliation{Bartol Research Institute and Dept. of Physics and Astronomy, University of Delaware, Newark, DE 19716, USA}

\author[0000-0001-9394-0007]{K. Andeen}
\affiliation{Department of Physics, Marquette University, Milwaukee, WI 53201, USA}

% ...

\date{\today}

\collaboration{432}{IceCube Collaboration}

\begin{abstract}
Recently, IceCube reported neutrino emission from the Seyfert galaxy NGC\,1068. Using 13.1 years of IceCube data, we present a follow-up search for neutrino sources in the northern sky. NGC\,1068 remains the most significant neutrino source among 110 preselected gamma-ray emitters while also being spatially compatible with the most significant location in the northern sky. Its energy spectrum is characterized by an unbroken power-law with spectral index $\gamma=3.4 \pm 0.2$. Consistent with previous results, the observed neutrino flux exceeds its gamma-ray counterpart by at least two orders of magnitude. 
Motivated by this disparity and the high X-ray luminosity of the source, we selected 47 X-ray bright Seyfert galaxies from the Swift/BAT spectroscopic survey that were not included in the list of gamma-ray emitters. When testing this collection for neutrino emission, we observe a 3.3\sigmas excess from an ensemble of 11 sources, with NGC~1068 excluded from the sample. Our results strengthen the evidence that X-ray bright cores of active galactic nuclei are neutrino emitters.

\end{abstract}


\section{Introduction}\label{sec:intro}
For over a decade, the IceCube Neutrino Observatory has been consistently detecting a diffuse flux of high-energy cosmic neutrinos \citep{IceCube:2013Science}.
While a fraction of this flux has recently been linked to the Galactic Plane \citep{IceCube:2023Science}, the majority remains isotropic, pointing to an extragalactic origin. Neutrinos are expected to be produced in interactions of protons with ambient matter and/or radiation within their cosmic sources. These interactions generate charged and neutral pions, which subsequently decay to produce neutrinos and gamma rays. While gamma rays can also arise from purely leptonic processes, neutrinos are only produced in hadronic interactions and are expected to be accompanied by gamma-ray emission \citep[e.g., and references therein]{neutrino_crs_connection:HalzenHooper_2002}. In 2017, IceCube detected a high-energy neutrino with a high probability of being of astrophysical origin from the direction of the blazar TXS\,0506+056. At the time of the neutrino arrival, the source exhibited enhanced gamma-ray activity. The chance probability of such an association was excluded at the 3\sigmas level, making TXS\,0506+056 the first candidate non-stellar astrophysical neutrino source \citep{IceCube:2018ScienceAlert} and supporting the theoretically anticipated correlation between neutrino and gamma-ray emissions (see \citealt{Ahlers:2018PrPNP} and references therein). 

Beyond the observation of TXS\,0506+056, IceCube has reported 4.2\sigmas evidence for TeV neutrino emission from the nearby active galactic nucleus (AGN) NGC\,1068 \citep{IceCube:2022Science}, notably without corresponding gamma-ray emission at similar energies. In fact, the GeV gamma-ray emission detected by {\it Fermi}-LAT \citep{1FGL} from NGC\,1068 is likely dominated by star-formation activity \citep{Eichmann:2022ApJ}, and its photon flux is lower than the observed neutrino flux. 
Its neutrino emission is likely produced in the immediate vicinity of the supermassive black hole (SMBH) that powers the AGN, most plausibly within the AGN's \textit{corona} \citep{Inoue:2019ApJ,Murase:2020PhRvL,Murase:2022ApJL,Padovani:2024NatAstro}. The corona -- a plasma of extremely hot electrons (\( \sim 10^9~\mathrm{K} \)) -- is responsible for the characteristic X-ray emission of AGN \citep[e.g.,][and references therein]{Padovani:2017AApR}. In this environment, ultraviolet photons are Compton up-scattered by hot electrons to keV energies, producing X-rays \citep{Liang:1979ApJ}. These coronal X-ray photons serve as effective targets for photomeson production when interacting with protons of energies around \( 100~\mathrm{TeV} \), leading to the generation of \( 1\text{--}10~\mathrm{TeV} \) neutrinos as observed by IceCube \citep{Padovani:2024NatAstro}. Remarkably, NGC\,1068 stands out as one of the X-ray-brightest AGN in the sky \citep{Marinucci:2016MNRAS,Ricci:2017ApJS}.

Building on the evidence of neutrino emission from NGC\,1068 and its interpretation as originating in the AGN corona, we present the results of dedicated neutrino searches aimed at further investigating this phenomenon.
In \autoref{sec:dataset}, we describe the dataset, obtained by extending the one used in \citet{IceCube:2022Science} with $\sim50\%$ more statistics, resulting in a total of 13.1 years of \(\nu_{\mu}\)-induced events from the northern sky.
In \autoref{sec:method} we introduce the analysis framework and the performed analyses, and in \autoref{sec:results} we present the results.
A survey of neutrino emission in the Northern Hemisphere reveals that the most significant excess remains spatially consistent with the position of NGC\,1068.
In addition, we examine a legacy list of 110 gamma-ray sources selected from the fourth Fermi Large Area Telescope catalog (4FGL-DR2 \citealt{4FGL:2020ApJS}), testing for neutrino emission from both individual objects and the list as a whole. Motivated by the gamma-ray-obscured and X-ray-bright nature of NGC\,1068, we also compile a new, targeted sample of 47 AGN selected among the intrinsically brightest objects in the BAT AGN Spectroscopic Survey (BASS) catalog \citep{Koss:2022ApJS}, and perform dedicated searches for neutrino emission from this population. The findings of the analyses are discussed in  \autoref{sec:discussion}.

A previous search by the IceCube Collaboration \citep{Icecube:2024Seyfert} investigated potential neutrino emission from a list of Seyfert galaxies, motivated by the neutrino excess from NGC\,1068. That study examined 27 Seyfert galaxies selected as the intrinsically brightest in the 2--10~keV band from the BASS catalog but did not find a statistically significant neutrino signal. This work updates the candidate source selection, focusing on X-ray AGN in the same catalog that are especially bright in the 20--50~keV band. The harder X-ray emission is more robust against both line-of-sight absorption and spectral features associated with low-temperature AGN coronae. Despite substantial overlap with the previous source list, the newly adopted selection criteria and increased statistics provide evidence for neutrino emission from a population of X-ray bright AGN in the northern sky.

\section{Neutrino Dataset}\label{sec:dataset}
The IceCube Neutrino Observatory is comprised of 86 strings, each instrumented with 60 digital optical modules (DOMs). The DOMs record the Cherenkov light produced by charged particles travelling through the Antarctic ice and are deployed at depths between 1450\,m and 2450\,m \citep{Icecube:2016Detector}. This full configuration, referred to as IC86, has been in continuous operation since May 13, 2011. Prior to that, from June 1, 2010, IceCube operated with 79 deployed strings (IC79), a near-complete geometry missing only seven strings. Of these, two belong to the DeepCore sub-array \citep{Icecube:2016Detector}, while the remaining five form an outer line on one edge of the array (see \autoref{fig:ICgeometry}).  
The IC79 dataset, comprising approximately 312 days of livetime, was reprocessed for this work using improved calibration and event filtering \citep{Icecube:2021PhRvD}, ensuring consistency with the IC86 data. The total dataset used in this study corresponds to 13.1 years of data, combining 0.9 years from IC79 and 12.2 years from IC86, collected up to November 28, 2023, and totaling approximately one million events.

\begin{figure}
    \centering
    \includegraphics[width=0.85\linewidth]{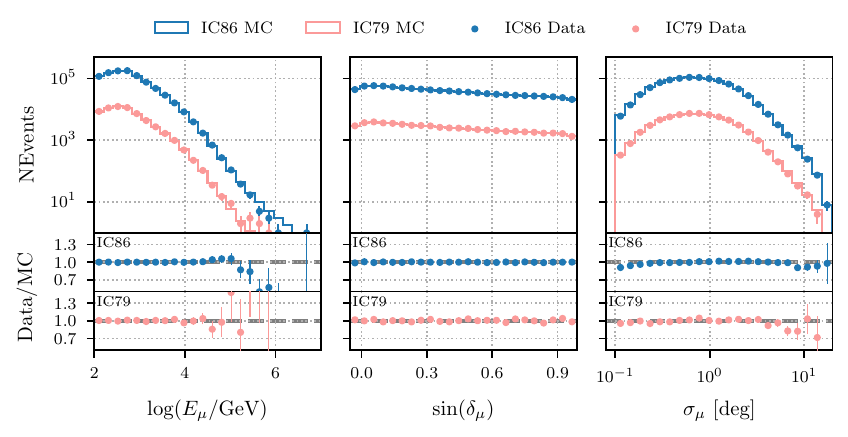}
    \caption{Agreement between the distributions of the analysis observables for experimental data (dots) and simulations (solid lines) for both IC79 (pink) and IC86 (blue). From left to right, we show the reconstructed muon energy, $E_{\mu}$, which ranges from 100~GeV to a few PeV, the sine of the reconstructed declination, sin($\delta_{\mu}$), and the reconstruction quality estimator on the muon track direction $\sigma_{\mu}$. In the lower panels, we display the ratios between the experimental data and the simulations for both detector configurations. The simulations show the sum of the atmospheric and the diffuse astrophysical neutrino flux components. The atmospheric neutrino flux assumes the GST model \citep{GST:2013FrPhy} for the primary cosmic-ray spectrum and \textit{Sybill2.3c} as interaction model \citep{Sybill2.3c}. The diffuse astrophysical component assumes a single power-law with spectral index $\gamma_{\mathrm{astro}} = 2.37$ and normalization $\phi_{\mathrm{astro}} = 1.44\times10^{-18}~\mathrm{GeV}^{-1}~\mathrm{cm}^{-2}~\mathrm{s}^{-1}~\mathrm{sr}^{-1}$ \citep{IceCube:2022NTDiffuse}.
    }
    \label{fig:dataMC}
\end{figure}

In this work, we focus on muons produced in charged-current interactions of muon neutrinos. As they traverse the detector, these muons emit Cherenkov light, creating track-like signatures. \textit{Tracks} are characterized by sub-degree angular resolution at energies above $\sim1$\,TeV (see \autoref{app:performance}), which makes this signature optimal for astrophysical source searches. We select events reconstructed with a declination ($\delta$) between $-5^\circ$ and $90^\circ$, encompassing the analyzed sky region ($-3^\circ < \delta < 81^\circ$). By focusing on the northern sky, the selection effectively suppresses atmospheric muon contamination, as such muons are absorbed when crossing the Earth before reaching the detector.
Overall, this selection yields a neutrino purity of 99.8\% \citep{IceCube:2022NTDiffuse}. The selected $\nu_\mu$-induced sample is predominantly composed of muons produced by atmospheric neutrinos and a subdominant unresolved diffuse astrophysical component, which contributes $\sim0.4\%$ to the total statistics of the sample, assuming the astrophysical flux measured by \cite{IceCube:2022NTDiffuse}. Both are treated as backgrounds to the search for point-like neutrino emission.
For each detected muon, we reconstruct the analysis observables: the muon track direction, $\boldsymbol{d}_{\mu}$, energy, $E_{\mu}$, and angular uncertainty, $\sigma_{\mu}$.
The 13-year dataset used in this analysis was processed following the same procedures as in \citet{IceCube:2022Science, Icecube:2024Seyfert}. While including more IC86 data is straightforward, incorporating IC79 data requires additional care due to the slightly different detector geometry.  
IC79 data and Monte Carlo simulations (MC) were reprocessed with the same reconstruction algorithms as IC86, and machine-learning models \citep{IceCube:2022Science} trained on IC86 simulations were successfully applied to IC79 events, achieving consistent performance across both configurations \citep{IC79_method:2023ICRC}.
In \autoref{fig:dataMC}, we show the agreement between the experimental data and the simulations for all relevant observables used in the analysis for both IC79 and IC86 configurations. Across all considered cases, the level of agreement is excellent, with maximal discrepancies of only a few percent in the high-statistics regions.

\section{Analysis Method}\label{sec:method}

In the search for neutrino point sources, we employ an unbinned maximum likelihood method, along with likelihood ratio hypothesis testing, as described in \citealt{BRAUN2008PSLLH}. The background-only hypothesis consists purely of contributions from the atmospheric and astrophysical diffuse emissions, while the signal hypothesis assumes an additional accumulation of astrophysical neutrinos clustered around a point-like source. The likelihood function for the total number of events in the sample $N$ is defined as:
\begin{equation}
    \label{eq:final_ps_llh}
    \mathcal{L}\left(n_\mathrm{s},\,\gamma,\,\boldsymbol{d}_{\mathrm{src}} \mid \boldsymbol{x} \right) =\prod_{i=1}^N\left\{\frac{n_\mathrm{s}}{N} \cdot  f_{\mathrm{S}}\left(\boldsymbol{x}_i \mid \gamma,\,\boldsymbol{d}_{\mathrm{src}}\right)+\left(1-\frac{n_\mathrm{s}}{N}\right) \cdot f_{\mathrm{B}}\left(\boldsymbol{x}_i\right)\right\}
    ,
\end{equation}
where $f_\mathrm{S}$ and $f_\mathrm{B}$ are the signal and background probability density functions (pdfs), respectively.
These pdfs are joint distributions that can be factorized into an energy and a spatial term, which together define the probability that a given event arises from a signal source or from background. An event $\boldsymbol{x}_i$ is defined by the vector of observables $(E_{\mu,i},\,\boldsymbol{d}_{\mu,i},\,\sigma_{\mu,i})$ (defined in \autoref{sec:dataset}),
$\boldsymbol{d}_{\mathrm{src}}$ denotes the direction of the analyzed source location in the sky, $n_\mathrm{s}$ is the mean number of signal events, and $\gamma$ is the spectral index of an assumed unbroken power-law spectrum of $\Phi=\Phi_0 \cdot (E/E_0)^{-\gamma}$, with neutrino energy $E$ and flux normalization $\Phi_0$ proportional to $n_\mathrm{s}$ at pivot energy $E_0=1$~TeV. The free parameters of the likelihood are the mean number of signal events, $n_\mathrm{s}$, and the spectral index, $\gamma$. 
The excellent agreement between data and simulations (see \autoref{fig:dataMC}) allows us to derive all observables' pdfs directly from the MC using the kernel density estimation (KDE) method \citep{KDE:Poluektov_2015}, as done in previous works \citep{IceCube:2022Science, Icecube:2024Seyfert}. For this analysis, we use updated KDE pdfs for IC86 -- now based on $\sim$2.5 times more simulated events, including muons from $\tau$ decays in $\nu_\tau$ interactions -- and construct them for IC79 for the first time. The two samples are then combined in the analysis by multiplying their likelihood functions, weighted according to their respective detection efficiencies.

The test statistic (TS) adopted in this work is the negative logarithm of the likelihood ratio between the background and the signal hypotheses, with the signal likelihood maximized over the two source parameters, $n_{\mathrm{s}}$ and $\gamma$:
\begin{equation}
    \label{eq:ts}
    \mathrm{TS} {\bf =} -2 \log\left(\frac{\mathcal{L}(n_\mathrm{s}=0 \mid \boldsymbol{x})}{\mathcal{L}(\hat{n}_\mathrm{s},\,\hat\gamma,\,\boldsymbol{d}_{\mathrm{src}} \mid\boldsymbol{x})}\right)=2\sum_i^N\log \left\{{\frac{\hat{n}_{\mathrm{s}}}{N}}\left(\frac{f_{\mathrm{S}}(\boldsymbol{x}_i \mid \hat\gamma,\,\boldsymbol{d}_{\mathrm{src}})}{f_{\mathrm{B}}(\boldsymbol{x}_i)}-1\right)+1\right\},
\end{equation}
where $\hat{n}_{\mathrm{s}}$ and $\hat\gamma$ are the parameter values which maximize the likelihood. The last equality in \autoref{eq:ts} uses \autoref{eq:final_ps_llh} for the signal and background cases and highlights the dependence of the TS on the ratio of the signal and background pdfs, $f_{\mathrm{S}}(\boldsymbol{x}_i \mid \hat\gamma,\,\boldsymbol{d}_{\mathrm{src}})/f_{\mathrm{B}}(\boldsymbol{x}_i)$, also referred to as $S/B$. The TS is used to assess the significance of the signal hypothesis relative to the background hypothesis: higher TS values indicate that the observed data is less compatible with the background-only hypothesis. The likelihood optimization is performed limiting the parameters in the ranges $[0,\,1000]$ for $n_{\mathrm{s}}$ and $[0.6,\,4.4]$ for $\gamma$, unless the source hypothesis assumes another spectral shape (e.g., a fixed spectral index or a spectrum different from the power-law).

This approach is applied to all the results presented in \autoref{sec:results}. First, we report on an unbiased search for neutrino emission across the northern sky. This method consists of dividing the sky into a grid of points and testing each of them for astrophysical neutrino emission using the aforementioned maximum-likelihood-ratio method.
Next, we present results from a targeted search for sources within two predefined lists: one containing 110 gamma-ray emitters and the other comprising 47 X-ray bright AGN.
For each list, we perform two tests:
\begin{itemize}
    \item A {\it catalog search}, where we evaluate the significance of neutrino emission with respect to the background for each source individually.
    \item A {\it binomial test}, to identify ensembles of sources that may show weak individual signals but yield a significant collective excess. For each of the \( N \) sources, we compute the local $p$-value and sort them in ascending order. For each rank \( k = 1, \ldots, N \), we then calculate the probability of observing \( k \) or more sources with $p$-values below the \( k \)-th smallest value in the list in a background-only scenario. In mathematical terms, the binomial $p$-value is defined as 
    \begin{equation}
        p_{\mathrm{binom}} = \sum_{i=k}^{N}\binom{N}{i}p_k^i(1-p_k)^{N-i}.
        \end{equation}
    The outcome of the test consists of the number of sources $k$ that provide the most significant collective excess above the background, i.e., the smallest binomial $p$-value among all the tested cases \citep{ATLAS-PHYS-PUB-2020-025}. 
\end{itemize}

\section{Results}\label{sec:results}

\subsection{Survey of the Northern Sky}\label{sec:skyscan}
\begin{figure}
    \centering
    \includegraphics[width=0.6\linewidth]{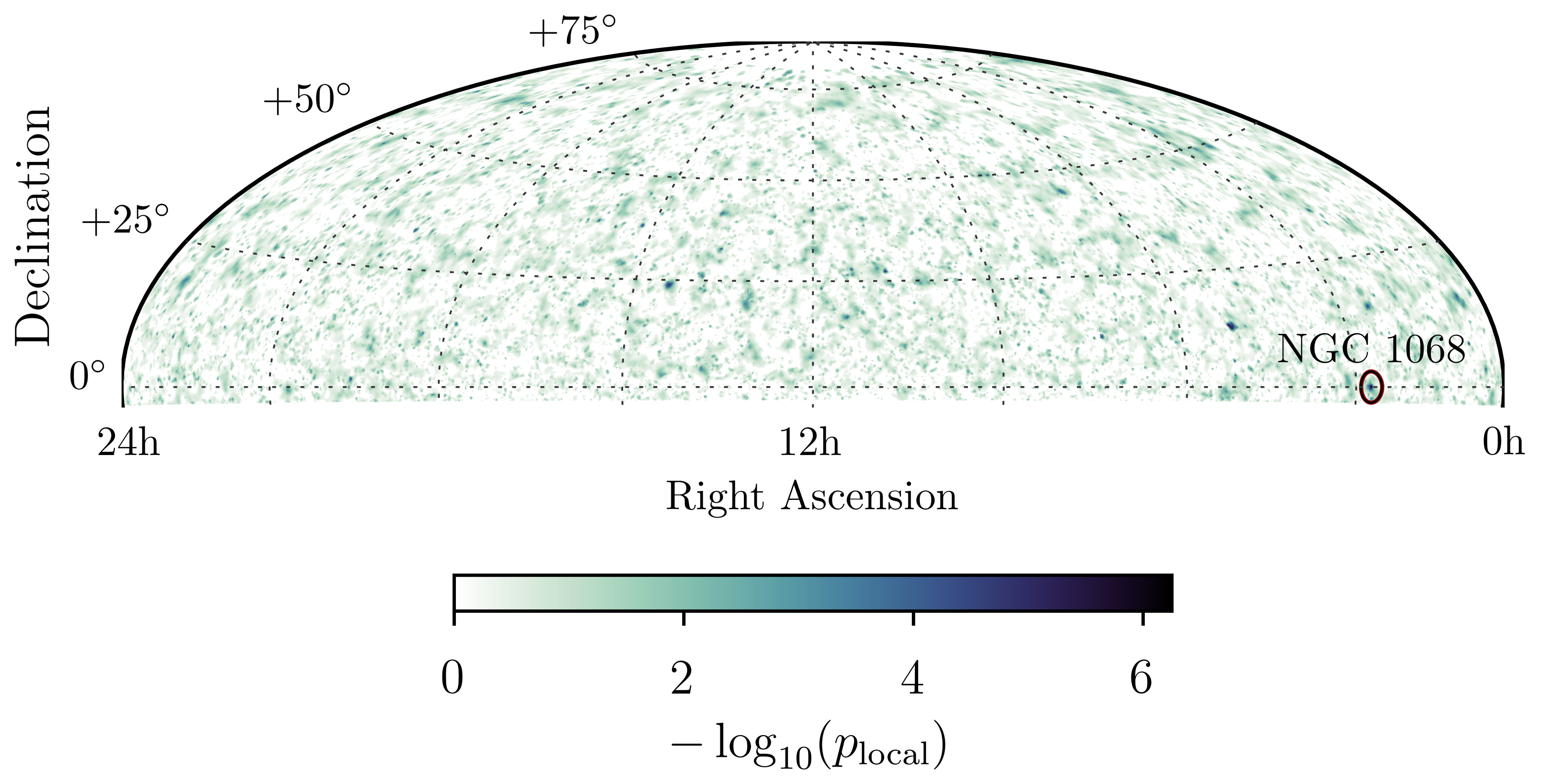}
    \includegraphics[width=0.3\linewidth]{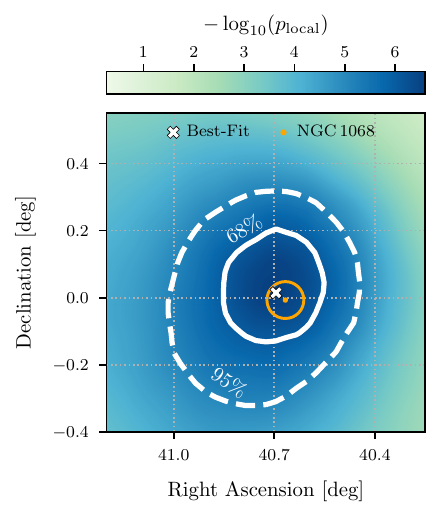}
    \caption{On the left, we show the $p$-value map of the northern sky, obtained under the hypothesis of a free spectral index. The map is shown in equatorial coordinates on a Hammer-Aitoff projection. The color bar represents the local significance of each pixel in the sky, and we highlight the location of the strongest emission, found to be spatially compatible with the Seyfert II galaxy NGC\,1068. On the right, we show a zoom-in of the hottest spot. The white cross represents the best-fit location, the solid (dashed) line represents the 68\% (95\%) uncertainty contour of the excess, and the orange dot and circle represent, respectively, the source location and its optical size \citep{Paturel:2003A&A}.}
    \label{fig:skymap}
\end{figure}

We performed an unbiased search for neutrino emission across the northern sky by dividing it into a grid of pixels of $\approx 0.052~\textrm{deg}^2$ using Healpix \citep{Gorski:2005ApJ} with resolution parameter \textsc{Nside} = 256. At each pixel, we optimized the likelihood ratio and extracted the best-fit values \(\hat{n}_\mathrm{s}\), \(\hat\gamma\), and the TS. The TS is then converted into a $p$-value by comparing it to the distribution of TS values from background-only (atmospheric and diffuse astrophysical neutrinos) simulations at the corresponding declination (see \autoref{app:significance} for details about the significance calculation).
To refine the localization of the most significant excess, we identified the 20 most significant hotspots\footnote{Defined as the lowest $p$-value pixel within a \(1^\circ\) radius.} and scanned these regions using a finer grid of pixels of $\approx 0.003~\textrm{deg}^2$ obtained by increasing the \textsc{Nside} resolution parameter to 2048. The lowest $p$-value from this refined search defines the final hottest spot in the sky.
As in the previous analysis \citep{IceCube:2022Science}, this search was performed under three spectral assumptions: with \(\gamma\) free, and fixed at \(\gamma = 2.0\) and \(\gamma = 2.5\). We report here the most significant excess out of the three: the one obtained for the free spectral index search. The other two result in most significant excesses at different locations, as detailed in \autoref{app:more-skyscans}. \autoref{fig:skymap} shows the $p$-value map under the free \(\gamma\) hypothesis, with the coordinates and best-fit parameters of the most significant excess listed in \autoref{tab:results}.

Consistent with \cite{IceCube:2022Science}, the analysis confirms that the most significant pixel lies within the optical extent of NGC\,1068, at an angular offset of \(0.04^\circ\) from its optical center (see \autoref{fig:skymap}, right panel).
The excess reaches a local significance of 5.0\sigmas, which corresponds to a global significance of 1.4\sigmas after accounting for the look-elsewhere effect from scanning the entire observable northern sky and testing three spectral hypotheses.
The location of the most significant spot remains fully compatible with NGC\,1068, and the best-fit point now aligns even more closely with the source than in the previous result, which placed it \(0.11^\circ\) away \citep{IceCube:2022Science}. However, the spatial likelihood contours have slightly widened due to the relatively large angular uncertainty of several newly detected low-energy events (with reconstructed energies mostly between 100\,GeV and 3\,TeV) associated with the source. The new events populate a part of the spectrum where the atmospheric background is more prominent and, together with statistical fluctuations, may also explain the reduction in global significance from 2.0\sigmas \citep{IceCube:2022Science} to 1.4\sigmas.
Despite this, the results remain consistent with the expected evolution of a steady neutrino emitter scenario for NGC\,1068.

\subsection{Search from a List of Gamma-ray Emitters}\label{sec:gamma-ray}

All-sky searches suffer from a large look-elsewhere effect due to the vast number of independent tests. To mitigate this penalty factor and improve sensitivity to weaker but persistent sources, we perform complementary searches on predefined lists of candidate emitters. Specifically, we run a catalog search and a binomial test on the legacy list of 110 gamma-ray sources, as employed in \citet{IceCube:2022Science}. 
Using 13.1 years of data, we confirm NGC\,1068 as the most significant source in the list, now with a global significance of 4.0\sigmas, after accounting for having tested 110 candidate sources. The best-fit spectrum has a flux normalization at 1~TeV of \(\hat\phi_0 = 4.7^{+1.1}_{-1.3} \times 10^{-11}~\mathrm{TeV}^{-1}~\mathrm{cm}^{-2}~\mathrm{s}^{-1}\) and a spectral index \(\hat\gamma = 3.4\pm0.2\)\footnote{The 1\sigmas statistical uncertainties on each flux parameter are derived from the one-dimensional profile likelihoods fixing the other parameter to its best-fit value.}.

\begin{figure}
    \centering
    \includegraphics[width=0.65\linewidth]{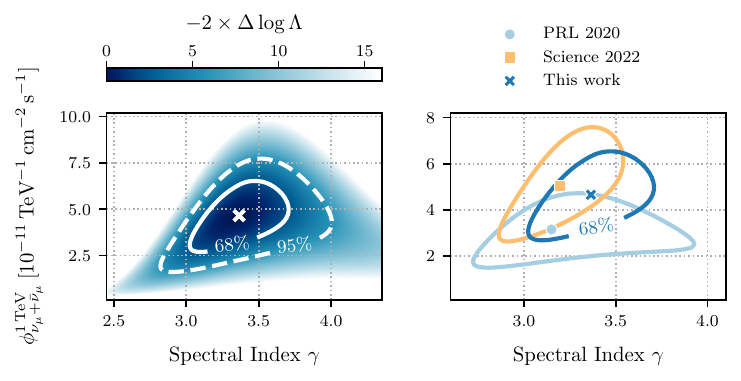}
    \caption{Left: Profile likelihood scan for the flux parameters of NGC\,1068. We highlight in white the best-fit result (cross) and the 68\% and 95\% uncertainty contours (solid and dashed lines), derived from Wilks' theorem \citep{wilks}. The color scale indicates the log-likelihood ratio difference between any point in the parameter space and the overall best-fitting point. Right: best-fit and 68\% contour comparison between this work (blue), \citealt{IceCube:2022Science} (yellow), and \citealt{IceCube:2020PhRvL} (light blue) -- the first IceCube analysis that found NGC\,1068 as the most significant source in a list, with a global significance of 2.9\sigmas. All contours include only statistical uncertainties.}
    \label{fig:llh-scan}
\end{figure}

As previously noted, the global significance has slightly decreased compared to \citet{IceCube:2022Science}, primarily due to a shift toward lower neutrino energies, where the atmospheric background is more prominent. As a result, the best-fit spectral index has slightly softened, increasing from \(\hat\gamma = 3.2\) to \(3.4\) (see \autoref{fig:llh-scan}). However, the two spectral indices remain largely compatible within their uncertainties (see \autoref{fig:NGC_energy} for a comparison of the best-fit spectrum measured in this work with the one previously reported in \citealt{IceCube:2022Science}). The fitted mean number of signal events has increased from \(\hat{n}_\mathrm{s} = 79\) to 102 (both affected by a 1\sigmas statistical uncertainty of $\sim 25\%$ as derived from the likelihood contours), consistent with a steady emission scenario.

To complement the single source search, we also perform a binomial test on the same list, following the approach in \citet{IceCube:2022Science}. As in the all-sky analysis, the test was conducted for three spectral assumptions: free \(\gamma\), \(\gamma = 2.0\), and \(\gamma = 2.5\). Here, we focus on the free-\(\gamma\) case, which yields the most significant result (see \autoref{app:more-binomials} for the others).
The most significant excess is found for 3 out of 110 sources: NGC\,1068, PKS\,1424+240, and TXS\,0506+056, in ascending order of $p$-value (see left panel of \autoref{fig:binomials}), consistent with previous results \citep{IceCube:2022Science}. Their fit parameters and local $p$-values are listed in \autoref{tab:results}. The global significance of the excess is 3.0\sigmas, slightly reduced compared to the 3.4\sigmas reported in \citet{IceCube:2022Science}. This decrease arises because the time-integrated significance of TXS\,0506+056 diminishes as additional data increase the background without new signal events, consistent with its reported time-variable behavior \citep{IceCube:2018ScienceAlert, IceCube:2018ScienceFlare}. Surprisingly, no additional gamma-ray sources from the list contribute significantly to the binomial excess, despite the increased exposure, highlighting the continued lack of broad correlation between gamma-ray brightness and neutrino emission in the current sample.

\subsection{Search from a List of X-ray-bright Active Galactic Nuclei}\label{sec:x-ray}

The emergence of NGC\,1068 as a neutrino source motivated the dedicated follow-up presented here. As one of the closest and brightest Seyfert galaxies, this association suggested that other AGN with similar properties might also emit neutrinos. A first attempt in this direction was made in \citet{Icecube:2024Seyfert}, which reported a 2.7\sigmas binomial excess from a set of 27 X-ray bright Seyfert galaxies selected from the BASS catalog.

Building on this, we constructed an updated list of nearby X-ray bright AGN as candidate neutrino sources. We selected sources classified as Seyfert galaxies in the BASS catalog that exhibit intrinsic hard X-ray fluxes between 20 and 50 keV of at least 20\% of that of NGC 1068. Although Seyfert galaxies are typically radio-quiet, the BASS classification is based solely on optical spectroscopy, and some nearby radio-loud galaxies may therefore appear in the sample. Since our focus is the correlation between neutrino production and X-ray brightness, such contamination is not a concern. The energy band of 20--50~keV was chosen to ensure robustness against line-of-sight absorption due to obscuring material in the circumnuclear environment. Absorption becomes negligible above $\sim10$~keV for column densities up to $\log N_{\mathrm{H}} \approx 23.5$ \citep{Koss:2022ApJS}, making this range a more reliable proxy for intrinsic source power than the softer 2–10~keV band used in \cite{Icecube:2024Seyfert}.
Using intrinsic fluxes up to 50~keV also minimizes potential bias against AGN with low-temperature coronae, whose spectra can cut off below 100~keV \citep{Fabian:2017MNRAS}, making them appear fainter in broader bands such as 14--195~keV. While the photon energies relevant for hadronic neutrino production are expected to lie in the softer $\sim1\text{--}10$~keV range, that band is highly susceptible to absorption. The 20--50~keV range therefore serves as a practical compromise: it is high enough to ensure robustness against absorption but not so high as to exclude plausible neutrino sources due to coronal spectral features. The resulting sample comprises 47 galaxies, excluding NGC\,1068 itself, and includes both Seyfert I and II types, thereby spanning a range of obscuration levels and ensuring a representative set of potential neutrino-emitting environments.

The selected X-ray bright, non-blazar AGN were tested under two spectral assumptions: a power-law with free spectral index \(\gamma\) and source-specific spectra from the core–corona model of \citet{Kheirandish:2021ApJ}, as in \citet{Icecube:2024Seyfert}. Here, we report only the results for the power-law case, which resulted in the highest statistical significance (see \autoref{app:more-results} for the core–corona model results).

\begin{figure}
    \centering
    \includegraphics[width=0.45\linewidth]{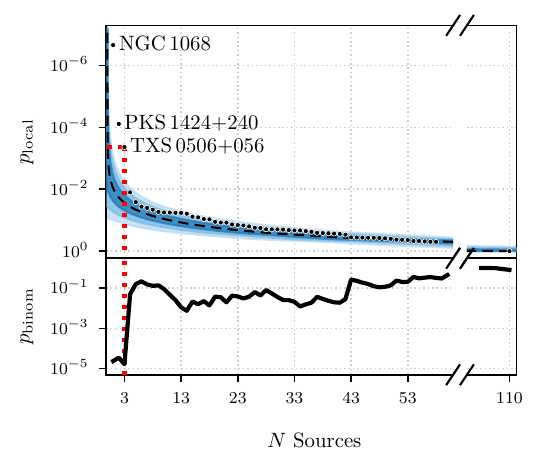}
    \includegraphics[width=0.45\linewidth]{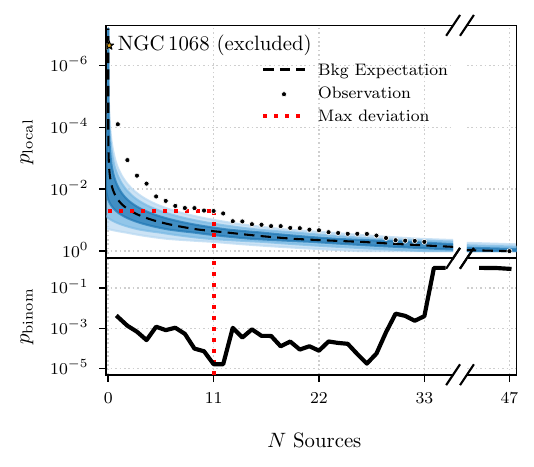}
    \caption{Visualization of the binomial test excess for the list of 110 gamma-ray emitters (left) and 47 X-ray-bright AGN (right). In the upper panels, we show the local $p$-values for each source, ordered from the lowest to the highest (black points), the background expectation (dashed black line), and its 1\sigmas, 2\sigmas, and 3\sigmas Poissonian uncertainties (shaded blue bands). In the lower panels, we report the binomial probability for each subset of sources. The most significant excess is highlighted by dotted red lines. To preserve readability, the plot is truncated at the point where sources reach a p-value of 1.0, beyond which the binomial probability is also equal to 1.0.}
    \label{fig:binomials}
\end{figure}

The most significant individual source, excluding NGC\,1068, is NGC\,7469, with a local significance of 3.8\sigmas and a global significance of 2.4\sigmas, accounting for 47 tested sources and two spectral hypotheses. The likelihood fit for this source returned a very hard spectral index of $\hat\gamma = 1.9$ and the excess is fully dominated by two high-energy events, already identified by IceCube as likely astrophysical and issued as neutrino alerts: IC220424A\footnote{\url{https://gcn.nasa.gov/circulars/31942}} and IC230416A\footnote{\url{https://gcn.nasa.gov/circulars/33633}} (see Appendix \ref{subsec:topsources} for more details and \citet{Giacomo_NGC7569:ApJ2025} for an independent study on this neutrino association). Most interestingly, the binomial test on this list reveals a collective excess from 11 out of 47 sources, with a \textbf{local} significance of 4.2\sigmas and a \textbf{global} significance of 3.3\sigmas after accounting for having tested different $p$-value thresholds and two spectral hypotheses (see \autoref{app:more-results} for more details). Among the 11 contributing sources, each has a local $p$-value $p_{\mathrm{local}} \leq 6\%$, while Monte Carlo simulations show that background-only realizations typically yield about five such sources.
The contributing sources are listed in \autoref{tab:results} and shown in the right panel of \autoref{fig:binomials}.
Including NGC\,1068 in the list (yielding 48 sources) increases the excess to 12 sources with a local significance of 4.5\sigmas, but does not qualitatively alter the result.

\begin{table*}
\centering
\caption{\label{tab:results}
\textbf{Summary of the results.} We report the most significant results from all tests performed in this work: the northern sky survey, the catalog searches, and the associated binomial tests. The type of test is indicated in the left-most column, with the corresponding results shown to the right. For each source, we report the equatorial coordinates (J2000 equinox), and the likelihood search results: number of signal events $\hat{n}_{\mathrm{s}}$, spectral index $\hat\gamma$, local $-\log_{10} p_{\mathrm{local}}$ and global $-\log_{10} p_{\mathrm{global}}$ (after accounting for all performed trials) $p$-values with their corresponding significance in brackets. The sources highlighted with an asterisk ($^*$) were included in the \citet{Icecube:2024Seyfert} selection.}
\begin{tabular}{lllrrrrcc}

\toprule
&& & R.A. & Dec. & $\hat{n}_\mathrm{s}$ & $\hat\gamma$ & $-\log_{10} p_{\mathrm{local}}$ & $-\log_{10} p_{\mathrm{global}}$ \\
\midrule

\multicolumn{2}{l}{Northern Sky Survey}\rule{0pt}{4ex}  & Hottest Spot& 40.69 & 0.02 & 102.6 & 3.4 & 6.6\,(5.0\sigmas) & 1.1\,(1.4\sigmas) \\
\midrule
\multicolumn{2}{l}{110 gamma-ray sources} \rule{0pt}{4ex} \\
\midrule
& Most significant source & NGC\,1068 & 40.67 & $-$0.01 & 102.2 & 3.4 & 6.6\,(5.0\sigmas) & 4.5\,(4.0\sigmas) \\
& Binomial test & 3 Sources & & & & & 4.8\,(4.2\sigmas) & 2.9\,(3.0\sigmas) \\
\cmidrule{2-9}
& Other sources in binomial excess & PKS\,1424+240 & 216.76 & 23.80 & 96.3 & 3.6 & 4.1\,(3.8\sigmas) & $-$\\
&  & TXS\,0506+056 & 77.36 & 5.69 & 4.9 & 1.9 & 3.4\,(3.3\sigmas) & $-$\\
\multicolumn{2}{l}{47 X-ray bright AGN} \rule{0pt}{4ex} \\
\midrule
&Most significant source& NGC\,7469 & 345.82 & 8.87 & 5.5 & 1.9 & 4.1\,(3.8\sigmas)& 2.1\,(2.4\sigmas)  \\
& Binomial test & 11 Sources & & & & & 4.8\,(4.2\sigmas) & 3.3\,(3.3\sigmas) \\
\cmidrule{2-9}
&Other sources in binomial excess & NGC\,4151$^{*}$ & 182.64 & 39.41 & 27.6 & 2.7 & 2.9\,(3.1\sigmas) & $-$ \\
&& CGCG\,420-015$^{*}$ & 73.36 & 4.06 & 35.3 & 2.7 & 2.4\,(2.7\sigmas) & $-$ \\
&& Cygnus\,A$^{*}$ & 299.87 & 40.73 & 3.4 & 1.6 & 2.2\,(2.5\sigmas) & $-$ \\
&& LEDA\,166445 & 42.68 & 54.70 & 57.1 & 4.4 & 1.8\,(2.1\sigmas) & $-$ \\
&& NGC\,4992 & 197.27 & 11.63 & 27.3 & 2.9 & 1.6\,(2.0\sigmas) &$-$ \\
&& NGC\,1194$^{*}$ & 45.95 & $-$1.10 & 43.2 & 4.4 & 1.5\,(1.8\sigmas) &$-$ \\
&& Mrk\,1498 & 247.02 & 51.78 & 39.9 & 3.6 & 1.4\,(1.7\sigmas) &$-$ \\
&& MCG\,+4-48-2$^{*}$ & 307.15 & 25.73 & 36.7 & 3.2 & 1.4\,(1.7\sigmas) & $-$ \\
&& NGC\,3079 & 150.49 & 55.68 & 33.8 & 3.6 & 1.3\,(1.7\sigmas) &$-$ \\
&& Mrk\,417 & 162.38 & 22.96 & 4.4 & 2.0 & 1.3\,(1.6\sigmas)& $-$ \\

\bottomrule

\end{tabular}
\end{table*}

\section{Discussion}\label{sec:discussion}

Thanks to IceCube’s continuous data-taking and remarkable long-term stability, we now have a view of the neutrino sky with unprecedented sensitivity (see \autoref{app:performance}). This study builds on that progress, presenting updated results from 13.1 years of observations. While TXS 0506+056 remains the only case claimed by IceCube linking high-energy neutrinos to a gamma-ray source, our results provide new evidence supporting an association between neutrino emission and X-ray bright AGN.
Specifically, we find a population-level excess from a sample of 47 X-ray bright, non-blazar AGN. Most of these sources do not exhibit high-energy gamma-ray emission. The few exceptions include NGC\,1068, whose GeV emission is likely related to starburst activity \citep{NGCstarburst:2020ApJ}, and NGC\,4151, possibly associated with 0.1--100\,GeV gamma-ray emission at the 5.5\sigmas level, perhaps connected to its ultra-fast outflow (UFO) \citep{Peretti:2025JCAP}. The interpretation of NGC\,4151 is further complicated by the presence of nearby blazars \citep{Omeliukh:2025A&A, Peretti:2025JCAP}, but in either case --- whether the gamma rays originate in the UFO or in the blazars --- theoretical expectations indicate that the observed neutrino flux cannot be explained \citep{Padovani:2024NatAstro,Peretti:2025JCAP,Omeliukh:2025A&A}, strengthening the case for neutrino generation in the X-ray corona, where TeV gamma rays are expected to be absorbed.

\begin{figure}
    \centering
    \includegraphics[width=0.6\linewidth]{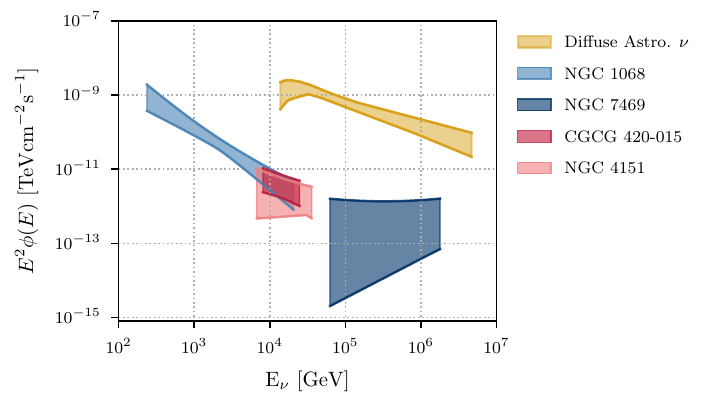}
    \caption{Best-fit neutrino all-flavor fluxes with their 1\sigmas uncertainty and constrained in the 95\% C.L. energy range of the top 4 sources in the list of X-ray bright AGN. The yellow band represents the most recent measurement of the diffuse astrophysical neutrino flux. The energy range is constrained at the 90\% C.L., while the flux normalization is shown with a 68\% C.L. \citep{abbasi2025:spectralbreak}.}
    \label{fig:spectra}
\end{figure}

Among the contributing AGN, we find both Seyfert I (e.g., NGC\,4151, NGC\,7469) and Seyfert II (e.g., NGC\,1068, CGCG\,420-015) galaxies, suggesting that the level of nuclear obscuration does not significantly impact the likelihood of neutrino emission.
In \autoref{fig:spectra}, we show the best-fit spectra for the four most significant X-ray bright AGN. Interestingly, the energy ranges of the contributing events vary across sources: NGC\,7469’s excess is dominated by two \(>100\,\mathrm{TeV}\) events, while NGC\,1068’s spectrum has shifted to lower energies, now constrained between 0.2 and 20.6\,TeV at 95\% C.L. (see \autoref{app:more-results} for more details on the energy range calculation), though still consistent with previous measurements.

The reduced significance of the NGC\,1068 signal -- despite the increased statistics -- could result from statistical fluctuations, spectral variability, or limitations of the assumed power-law model. An alternative explanation may be time variability \citep{Pranav:2022icrc}, which was not tested in this work due to the focus on steady emission scenarios. Future time-dependent studies will be essential to explore this possibility. Improved spectral constraints, particularly below 100\,GeV, will also be crucial for future observations.

Taken together, our findings suggest that X-ray bright AGN are promising contributors to the observed diffuse extragalactic neutrino flux. This builds on a growing body of independent results: a 2.7\sigmas binomial excess from NGC\,4151 and CGCG\,420-015 \citep{Icecube:2024Seyfert}, a 2.9\sigmas excess from NGC\,4151 in a hard X-ray AGN study \citep{Icecube2024:hardXray}, and a 3.0\sigmas signal from a stacking analysis of 14 Seyfert galaxies in the southern sky \citep{Southern_Seyfert:2025ICRC}.
While current significance levels remain marginal, the emerging picture points to a new class of potential neutrino sources: X-ray bright AGN. These results yield growing support for theoretical models of neutrino production in the coronal regions surrounding SMBHs.
At the same time, the very high-energy events associated with NGC\,7469 cannot be accommodated by the custom core-corona model \citep{Kheirandish:2021ApJ} tested in \cite{Icecube:2024Seyfert} and in this work (see \autoref{app:more-results}). This suggests that X-ray bright AGN may not all share the same cosmic-ray acceleration or neutrino production conditions, and that further measurements will be essential.
Future data from IceCube and next-generation detectors \citep{KM3NeT:2016JPhG,Gen2:2021JPhG,PONE:2020NatAs}, combined with multi-wavelength observations, will be critical to confirm this connection and further unravel the components of the diffuse astrophysical neutrino flux.

\section*{Acknowledgments}
The IceCube collaboration acknowledges the significant contributions to this manuscript from Chiara Bellenghi, Tomas Kontrimas and Elena Manao. The authors gratefully acknowledge the support from the following agencies and institutions:
USA {\textendash} U.S. National Science Foundation-Office of Polar Programs,
U.S. National Science Foundation-Physics Division,
U.S. National Science Foundation-EPSCoR,
U.S. National Science Foundation-Office of Advanced Cyberinfrastructure,
Wisconsin Alumni Research Foundation,
Center for High Throughput Computing (CHTC) at the University of Wisconsin{\textendash}Madison,
Open Science Grid (OSG),
Partnership to Advance Throughput Computing (PATh),
Advanced Cyberinfrastructure Coordination Ecosystem: Services {\&} Support (ACCESS),
Frontera computing project at the Texas Advanced Computing Center,
U.S. Department of Energy-National Energy Research Scientific Computing Center,
Particle astrophysics research computing center at the University of Maryland,
Institute for Cyber-Enabled Research at Michigan State University,
Astroparticle physics computational facility at Marquette University,
NVIDIA Corporation,
and Google Cloud Platform;
Belgium {\textendash} Funds for Scientific Research (FRS-FNRS and FWO),
FWO Odysseus and Big Science programmes,
and Belgian Federal Science Policy Office (Belspo);
Germany {\textendash} Bundesministerium f{\"u}r Bildung und Forschung (BMBF),
Deutsche Forschungsgemeinschaft (DFG),
Helmholtz Alliance for Astroparticle Physics (HAP),
Initiative and Networking Fund of the Helmholtz Association,
Deutsches Elektronen Synchrotron (DESY),
and High Performance Computing cluster of the RWTH Aachen;
Sweden {\textendash} Swedish Research Council,
Swedish Polar Research Secretariat,
Swedish National Infrastructure for Computing (SNIC),
and Knut and Alice Wallenberg Foundation;
European Union {\textendash} EGI Advanced Computing for research;
Australia {\textendash} Australian Research Council;
Canada {\textendash} Natural Sciences and Engineering Research Council of Canada,
Calcul Qu{\'e}bec, Compute Ontario, Canada Foundation for Innovation, WestGrid, and Digital Research Alliance of Canada;
Denmark {\textendash} Villum Fonden, Carlsberg Foundation, and European Commission;
New Zealand {\textendash} Marsden Fund;
Japan {\textendash} Japan Society for Promotion of Science (JSPS)
and Institute for Global Prominent Research (IGPR) of Chiba University;
Korea {\textendash} National Research Foundation of Korea (NRF);
Switzerland {\textendash} Swiss National Science Foundation (SNSF).

\bibliography{references}{}
\bibliographystyle{aasjournal}

\begin{appendix}
\section{IC86 and IC79 geometry comparison}\label{app:dataset}
The results presented in this letter were obtained using a dataset comprising data recorded with two different detector configurations, as explained in \autoref{sec:dataset}.

\begin{figure}
    \centering
    \includegraphics[width=0.3\linewidth]{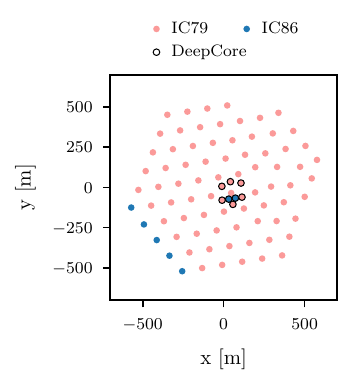}
    \caption{Disposition of the strings in the IceCube coordinate system. The blue dots mark the last seven strings added, which changed the detector from the IC79 to the IC86 configuration. For this work, only the five strings added to the edge of the detector are relevant, as the other two are part of the DeepCore detector (dots outlined in black), excluded from the analysis.}
    \label{fig:ICgeometry}
\end{figure}

In \autoref{fig:ICgeometry}, we show a top-view of the IceCube array, highlighting the two configurations with different colors. Apart from two strings belonging to the DeepCore array, the only difference between the IC86 and IC79 arrays consists of 5 strings on one edge of the detector. Due to the minimal changes in the overall geometry, data recorded in this configuration has minimal losses in terms of reconstruction quality and resolution compared to data recorded using the full IceCube array. In the left panel of \autoref{fig:KDEs}, we display the angular resolution of the events in the sample for both configurations. Both the medians and the central 1\sigmas quantile of the angular resolutions show minimal differences over the whole energy range, proving the comparable quality of IC79 and IC86 data.

\section{Analysis Performance}\label{app:performance}

\begin{figure}
    \centering
    \includegraphics[width=0.8\linewidth]{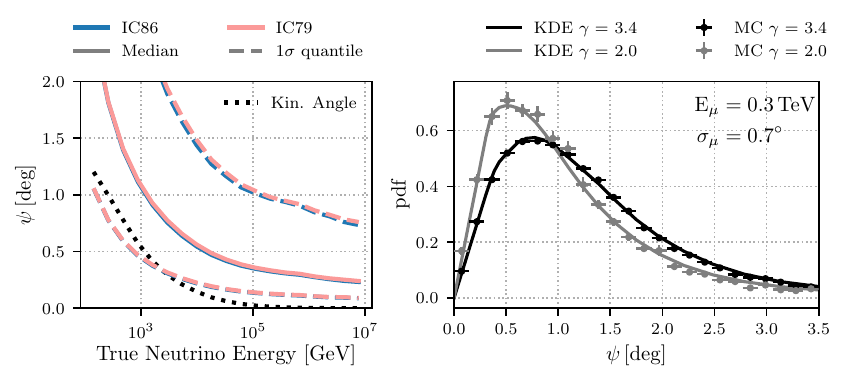}
    \caption{Left panel: Angular resolution for both IC86 (blue) and IC79 (pink) data in the neutrino energy range employed in the analysis. The solid line represents the median, while the dashed lines denote the central 1\sigmas quantiles (16\% and 84\% quantiles). The black dotted line marks the median kinematic angle, i.e., the median angular separation between the incoming neutrino and the muon produced in the interaction. Right panel: Two examples of KDE pdfs for two spectral assumptions (solid lines) superimposed with the MC data (dots). This example shows the pdfs for a muon energy of 300 GeV and an uncertainty on the angular reconstruction of 0.7$^{\circ}$.}
    \label{fig:KDEs}
\end{figure}

As described in \autoref{sec:method}, the likelihood analysis used in this work incorporates an improved description of the spatial and energy pdfs, which are parameterized from simulations using the Kernel Density Estimation (KDE) method \citep{KDE:Poluektov_2015}. The right panel of \autoref{fig:KDEs} shows two spatial pdfs obtained for different spectral indices.
Because the signal pdfs are derived directly from simulation, they accurately capture the relationship between reconstructed event properties and the underlying signal hypothesis. In particular, as seen in the right panel of \autoref{fig:KDEs}, events with the same reconstructed muon energy ($E_\mu$) and angular uncertainty ($\sigma_{\mu}$) can have different likelihoods of originating from a given source depending on their angular separation $\psi$ and the given spectral assumption. This effect arises because lower-energy neutrinos, which dominate for softer spectra, produce muons with larger kinematic angles from the parent particle (see the left panel of \autoref{fig:KDEs}), leading to a broader spatial distribution. Hence, incorporating the full simulation-based dependence of $\psi$ on energy and spectral shape improves our ability to accurately characterize the spectral emission of a source.

\begin{figure}
    \centering
    \includegraphics[width=0.8\linewidth]{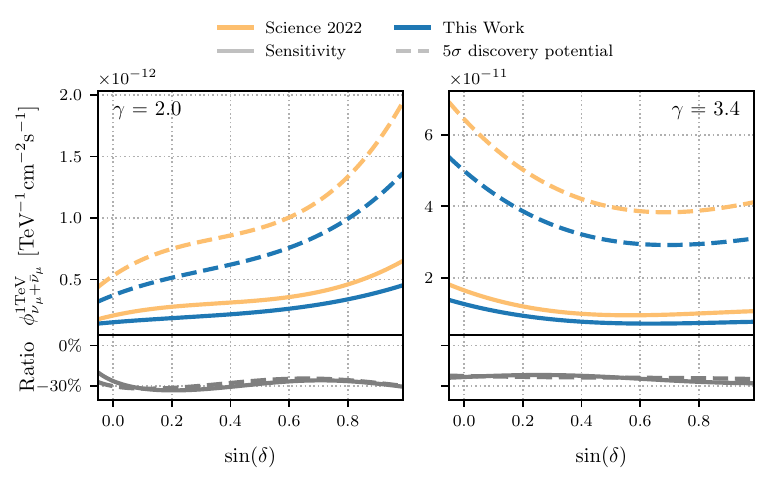}
    \caption{In the upper panels, we show the sensitivity (solid) and discovery potential (dashed) fluxes for a spectral index $\gamma = 2.0$ (left) and $\gamma = 3.4$ (right) for this analysis (blue) and \cite{IceCube:2022Science} (yellow). In the lower panels, we show the percentage ratio between the two. In both cases, the sensitivity and discovery potential improve by up to 30\% compared to previous results.}
    \label{fig:sens}
\end{figure}

In addition to the improved description of the pdfs used in the likelihood analysis, this work also benefits from a significant increase in available statistics. The dataset analyzed here spans over 13.1 years of uniformly processed IceCube data, resulting in a final event sample of approximately one million events, a $\sim$50\% increase compared to the \cite{IceCube:2022Science} analysis. 
To quantify the analysis capability of observing an astrophysical signal we use the \textit{sensitivity} and the 5\sigmas \textit{discovery potential} neutrino fluxes. The sensitivity is defined as the average flux a point source would need to exceed the median test-statistic (TS) value obtained from background-only simulations in 90\% of the cases. The 5\sigmas discovery potential is instead defined as the average flux needed to exceed the 5\sigmas quantile of the background TS distribution in 50\% of the cases. In \autoref{fig:sens} we show the sensitivity and the 5\sigmas discovery potential fluxes of this analysis, comparing them to those of \cite{IceCube:2022Science}. This work shows an increased sensitivity to astrophysical signals up to $\sim$30\% across the whole analyzed declination range for both hard ($\gamma = 2.0$) and soft ($\gamma = 3.4$) spectral indices.

\section{Significance Calculation for the Likelihood Ratio Test}\label{app:significance}
\begin{figure}
    \centering
    \includegraphics[width=0.5\linewidth]{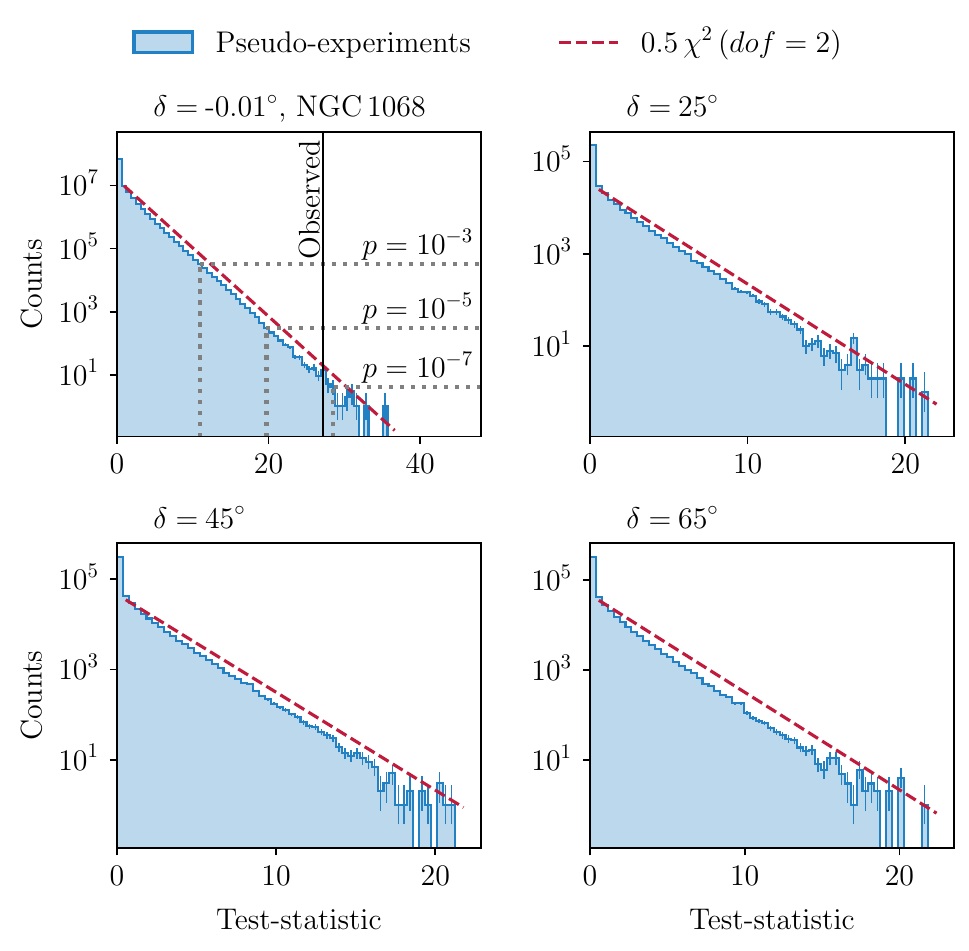}
    \caption{Test-statistic distributions under the null hypothesis at four representative declinations for the floating spectral index hypothesis. The blue histograms show the distributions obtained from background-only pseudo-experiments. The red dashed curve indicates the half-$\chi^2$ with 2 degrees of freedom ($dof$), shown only for reference. The upper left panel, in particular, shows the distribution of $\sim 10^8$ background-only TS values obtained at the location of NGC~1068 together with the observed TS (solid black line), and the TS values that would correspond to various local significance levels in terms of $p$-values at that location (dotted grey lines).}
    \label{fig:bkg_ts_examples}
\end{figure}

The calculation of the sensitivity, 5\sigmas discovery potential, or of the significance of an observed TS value requires the determination of the TS distribution under the null hypothesis.
Wilks’ theorem states that, under suitable regularity conditions, the likelihood-ratio test statistic asymptotically follows a chi-squared ($\chi^2$) distribution \citep{wilks}. However, in the analysis presented in this work (see \autoref{sec:method}), the signal-strength parameter lies on the boundary of the parameter space under the null hypothesis ($n_\mathrm{s} = 0$), thereby violating the required regularity conditions. In this situations, the asymptotic $\chi^2$ approximation is formally invalid \citep[see, e.g.,][]{protassov_2002}.
For this reason, no $\chi^2$ distribution is used anywhere in this work to compute local or global significances for rejecting the null hypothesis. Instead, the TS distribution under the null hypotheses is empirically determined using a large ($\sim10^6$ or more if needed) number of background-only pseudo-experiments. Background-only pseudo-datasets are generated by injecting events according to the expected atmospheric and astrophysical diffuse neutrino fluxes observed by IceCube. Each pseudo-dataset is then analyzed using the same likelihood-ratio maximization procedure applied to the experimental data.
\autoref{fig:bkg_ts_examples} shows the background TS distributions at four representative declinations (including that of NGC~1068). The red dashed curve indicates the half-$\chi^2$ distribution with two degrees of freedom, corresponding to the difference in the number of free parameters between the signal (2) and background (0) hypotheses, only for illustrative comparison; the clear discrepancies demonstrate explicitly that the $\chi^2$ approximation does not accurately describe the null distribution in this analysis.

The local $p$-value (i.e, prior to accounting for the look-elsewhere effect) for each tested sky direction is defined as the survival probability of the observed TS with respect to the empirically determined background-only TS distribution at that direction. An example is shown in the upper-left panel of \autoref{fig:bkg_ts_examples}, where the observed TS at the location of NGC~1068 (TS = 27.2) is compared to the TS distribution obtained from $\sim 10^8$ pseudo-experiments.

Global $p$-values (i.e., accounting for penalization due to the look-elsewhere effect) are likewise obtained from pseudo-experiments, in both the all-sky scan and the binomial test. This is necessary because, in those two cases, the tested spectral hypotheses are correlated, and any analytical corrections to $p$-values become either inaccurate or overly conservative. In each result subsection of \autoref{app:more-results}, we specify the exact procedure used to evaluate the final global significance of the result.

\section{Additional Results}\label{app:more-results}

In addition to the results presented in \autoref{sec:results}, we conducted several other tests that did not yield statistically significant outcomes and are therefore not included in the main body of the paper. Nevertheless, they have been taken into account to correct the global significance of our findings for the look-elsewhere effect. For completeness, we summarize these tests in the following subsections.

\subsection{All-sky Searches with Fixed Spectral Index}\label{app:more-skyscans}

Alongside the all-sky scan using a free spectral index we performed two supplementary tests in which the spectral index $\gamma$ was held fixed, as done in previous searches \citep{IceCube:2022Science}, and to be consistent in the calculation of the global significance accounting for the look-elsewhere effect. We tested values of $\gamma = 2.5$ and $\gamma = 2.0$. The former was chosen due to its similarity to the best-fit astrophysical neutrino flux \citep{globalfit_astro_pl:ICRC23}, while the latter reflects expectations for neutrino emission in scenarios governed by Fermi acceleration mechanisms.

The global significance of the result reported in \autoref{sec:skyscan} cannot be calculated analytically because the tested spectral hypotheses are highly correlated, as are adjacent tested locations in the sky. The correction is done in two steps: first, we account for testing multiple correlated locations in the sky by building the distribution of the minimum $p$-values from the pseudo-experiments for each spectral hypothesis. The fraction of realizations with smaller values than the observed $p$-values gives the scan-corrected significance. In the second step, the three scan-corrected $ p$-values (one per all-sky scan) are compared with the distribution of the minimum $p$-values obtained by analyzing the same pseudo-data under the three different signal hypotheses. The fraction of realizations resulting in a $p$-value smaller than the scan-corrected $p$-value yields the final globally-corrected $p$-value.

Both fixed-index scans revealed hotspot locations different from those found in the free-index case but with lower local significances. For the fixed $\gamma = 2$ case, the most significant hotspot was located at $(\alpha, \delta) = (77.01^\circ, 12.98^\circ)$, with a best-fit number of signal events of $\hat{n}_\mathrm{s} = 16.8$ and a local significance of approximately 4.9\sigmas. In the $\gamma = 2.5$ case, the hotspot appeared at $(\alpha, \delta) = (161.48^\circ, 27.32^\circ)$, with $\hat{n}_\mathrm{s} = 34.3$ and a local significance of around 4.5\sigmas. Neither of these hotspots is spatially compatible with any known source in the 4FGL-DR4 gamma-ray \citep{4FGLDR4:2023arXiv} or eRASS1 X-ray \citep{eRASS1:2024A&A} catalogs. However, we note that the hottest spot from the $\gamma=2.5$ scan is located $0.54^{\circ}$ away from the 4FGL source NVSS\,J104516+275136.

\begin{figure}
    \centering
    \includegraphics[width=0.4\linewidth]{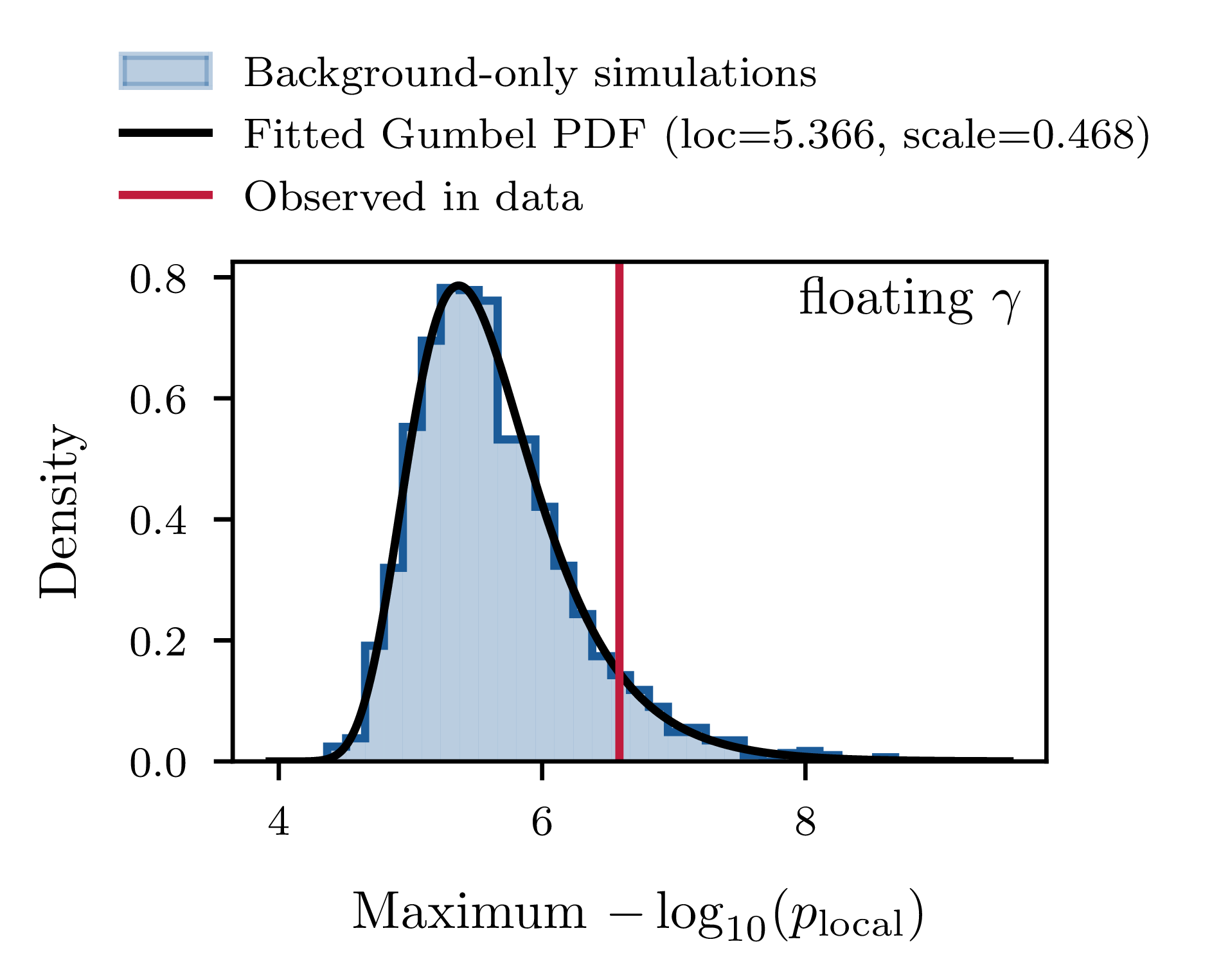}
    \caption{Distribution of the maximum local $-\log_{10}p$-value from 2000 background-only sky scan simulations under the floating spectral index hypothesis. The histogram shows the distribution of the most significant fluctuation per pseudo-experiment, while the black curve represents the best-fit Gumbel probability density function. The vertical red line indicates the maximum $-\log_{10}p$ observed in the experimental data. Equivalent agreement is found for the other spectral hypotheses.}
    \label{fig:gumbel}
\end{figure}

As a cross-check of the statistical behavior of the sky scans, we verified that the distribution of the most significant $p$-values obtained from 2000 background-only pseudo-experiments follows the expectation from Extreme Value Theory (EVT, \citealt{Coles2001}). For each spectral hypothesis, we constructed the distribution of the maximum $-\log_{10}(p)$ and fitted it with a Gumbel distribution, which is the limiting distribution for the maximum of a large number of independent random variables with exponentially decaying tails. An example fit for the floating spectral index hypothesis is shown in \autoref{fig:gumbel}. The Anderson--Darling goodness-of-fit test \citep{AndersonDarling1954} shows no evidence of deviations from the Gumbel hypothesis, with an $A^2$ statistic of 0.147 (the critical value to reject the Gumbel hypothesis at 95\% C.L. is 0.754). Similar agreement is found for the other two spectral hypotheses.
Scan-corrected $p$-values derived from the fitted Gumbel distributions are fully consistent with those obtained directly from the pseudo-experiments, confirming the expected asymptotic extreme-value behavior of the sky scan fluctuations. 
However, pseudo-experiments remain necessary for two reasons. First, the parameters of the Gumbel distribution cannot be determined analytically due to non-trivial correlations between adjacent pixels and must be calibrated from simulations. Second, the final global correction involves the minimum of only three strongly correlated, already scan-corrected $p$-values corresponding to the three tested spectral hypotheses, which does not lie in an asymptotic extreme-value regime and is not described by a Gumbel distribution. The final multi-hypothesis correction can therefore be obtained exclusively from pseudo-experiments, while the EVT analysis serves as a validation of the scan-correction procedure.

\subsection{Binomial tests with Fixed Spectral Shapes}\label{app:more-binomials}

To preserve consistency with the analysis presented in \cite{IceCube:2022Science}, we also performed the binomial test for the fixed spectral index cases using our list of 110 gamma-ray emitters. As with the all-sky scans, these tests did not yield statistically significant results. For $\gamma = 2$, we found 13 sources (including NGC~1068, TXS~0506+056, and PKS~1424+240) out of 110 with a local p-value of approximately 0.007, while for $\gamma = 2.5$, the test returned a local p-value of about 0.001 for two sources (NGC~1068 and TXS~0506+056). Since the local p-values of these excesses are orders of magnitude lower than those obtained when performing the test with a floating spectral index, we did not further investigate these results.

For the list of 47 X-ray bright AGN, we also applied the binomial test assuming the core–corona emission model \citep{Kheirandish:2021ApJ}. The core-corona model predicts the shape of neutrino flux for each tested source candidate based on the intrinsic X-ray luminosities as reported in the BASS catalog \citep{Koss:2022ApJS}, where additional model parameters are fixed to match the NGC~1068 neutrino flux \citep{Icecube:2024Seyfert}. The $f_\mathrm{S}$ signal pdfs in the likelihood function (\autoref{eq:final_ps_llh}) were constructed using the same KDE method as in the power-law case but assuming the predicted flux shape. It results in $n_{\mathrm{s}}$ as the only free parameter in the likelihood ratio test (\autoref{eq:ts}), that corresponds to the neutrino flux normalization. This is not the first time such a model has been tested using IceCube track-like events. \cite{Icecube:2024Seyfert} previously reported a 2.7\sigmas excess from a binomial test applied to a list of 27 Seyfert galaxies. Given the substantial overlap between the two samples, 23 out of 47 sources (or 24 out of 48 when including NGC\,1068), we applied the same model to our list. However, this analysis should not be interpreted as a direct follow-up to \cite{Icecube:2024Seyfert}. The selection criteria used to compile the source list presented here differ from previous ones as described in \autoref{sec:x-ray}.

The binomial test based on the core-corona model yielded a local p-value of approximately 0.001 for 3 sources out of 47. When NGC\,1068 is included in the sample, the p-value further decreases to $\sim10^{-5}$. Notably, two of the three sources contributing to the excess, NGC\,4151, and CGCG\,420-015, were also identified in \cite{Icecube:2024Seyfert}. Moreover, these sources also contribute to the excess observed under the power-law spectral assumption, further supporting the hypothesis that they may be genuine neutrino emitters.

As discussed before, the global significance of the results cannot be calculated analytically. Therefore, we compute them from pseudo-experiments. Similarly to what is described in \autoref{app:more-skyscans}, the a posteriori correction is obtained in two steps. First, we account for testing multiple significance thresholds by comparing the most significant excess with the distribution of the minimum $p$-values from background-only pseudo-experiments generated for each individual hypothesis. The fraction of realizations with smaller values than the observed result gives the threshold-corrected significance. In the second step, this threshold-corrected $p$-value is compared to the distribution of the minimum $p$-values obtained by analyzing the same pseudo-data under the different signal hypotheses. The fraction of realizations resulting in a $p$-value smaller than the corrected $p$-value yields the final globally-corrected $p$-value.

\subsection{Most significant sources}\label{subsec:topsources}
\begin{figure}
    \centering
    \includegraphics[width=0.3\linewidth]{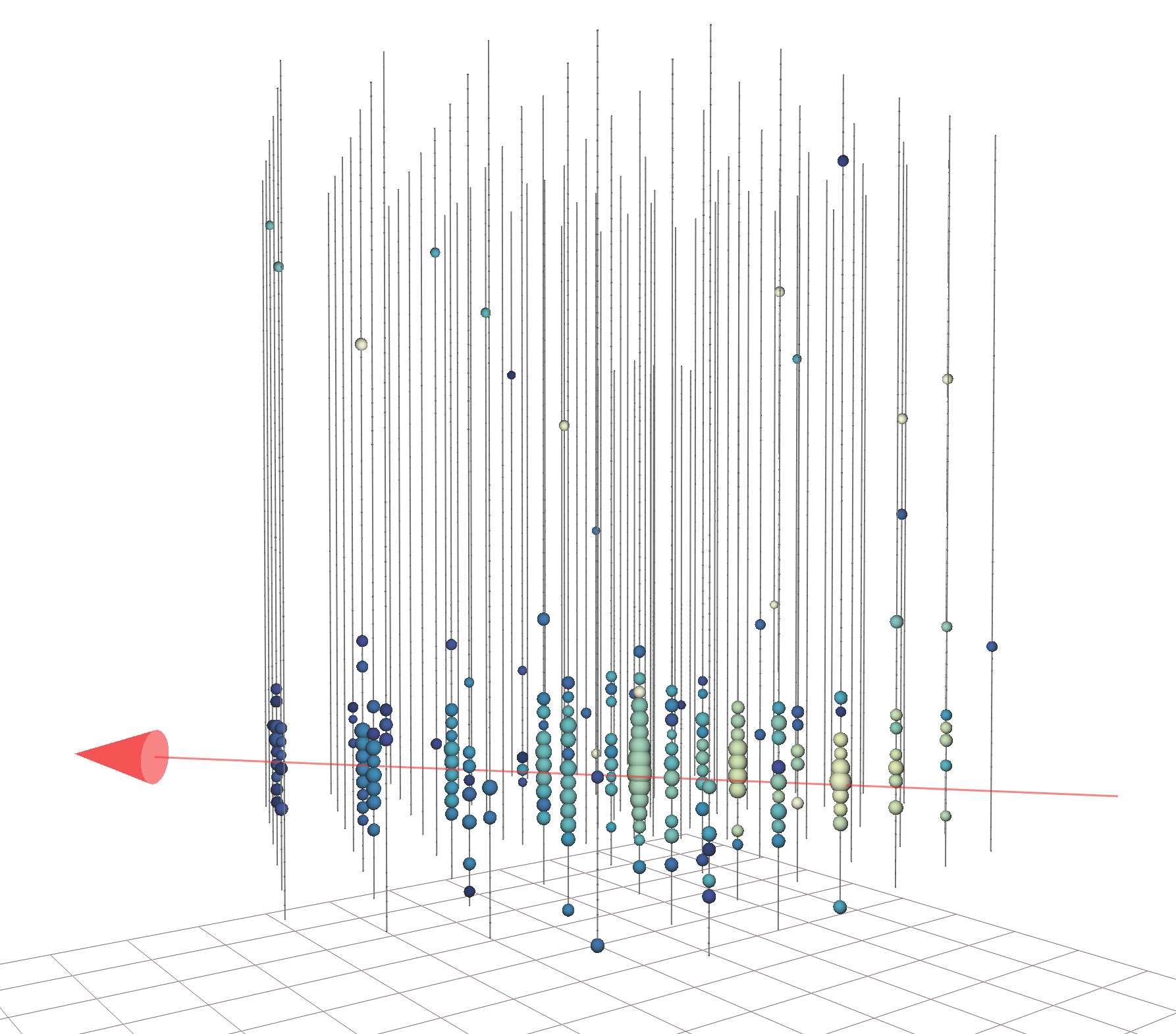}
    \includegraphics[width=0.3\linewidth]{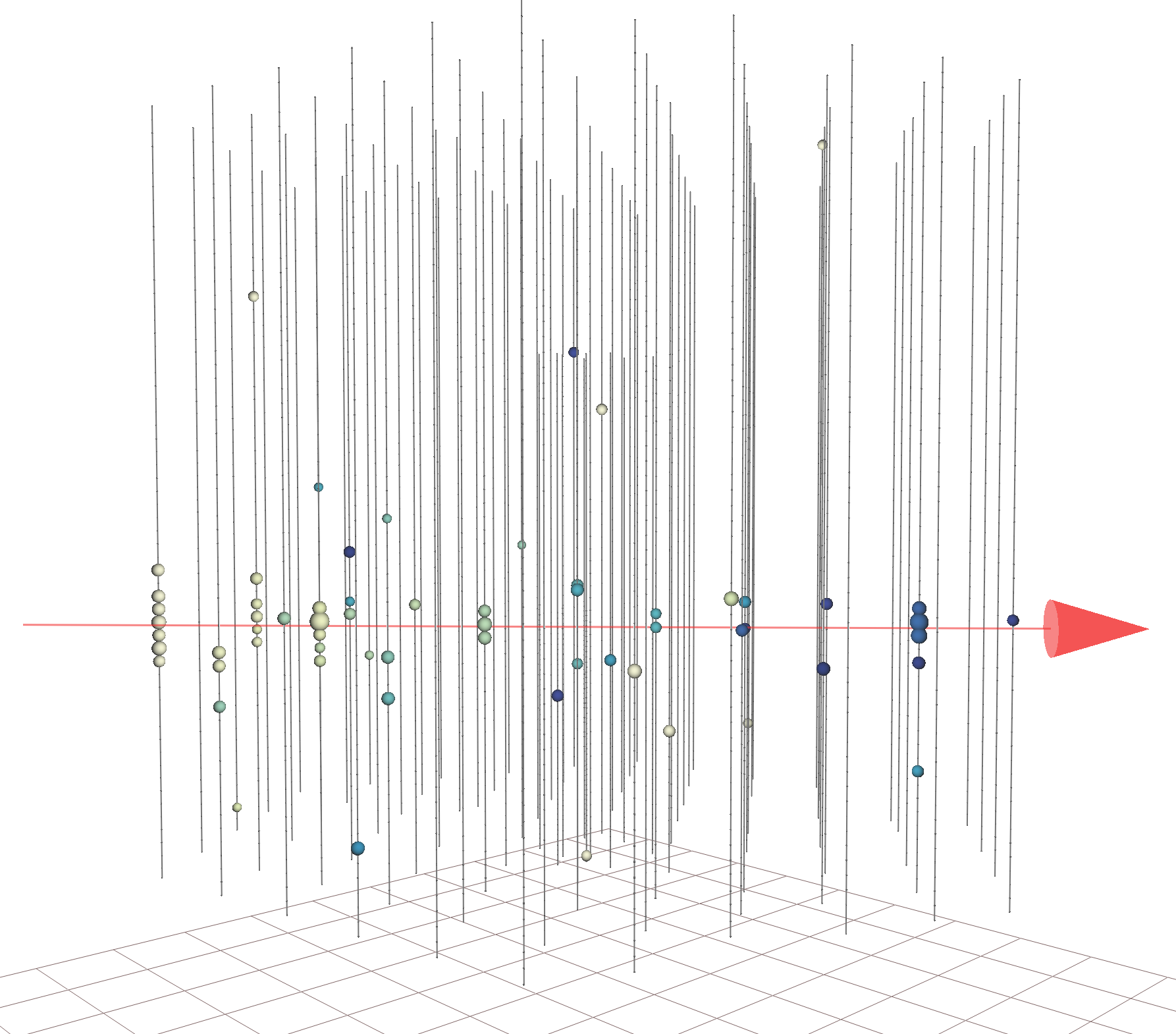}
    \includegraphics[width=0.3\linewidth]{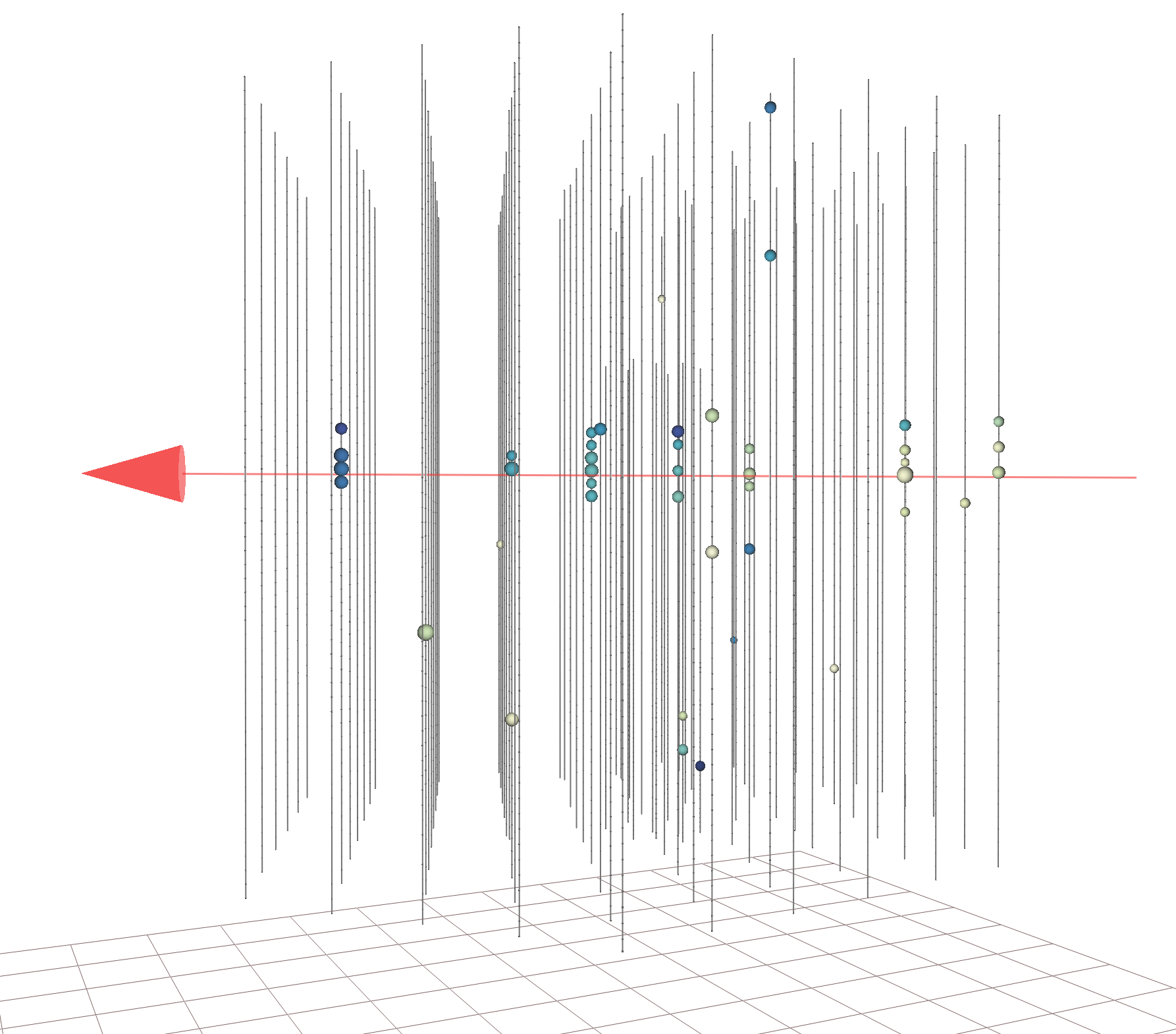}
    \caption{Event views of three of the events associated with the neutrino excess of NGC\,1068. The leftmost event is the one that contributes the highest $S/B$; in the central panel, we show the event that contributes the most among those added in this new iteration of the sample; in the rightmost panel, we show the highest contributing event from the IC79 season. The colored blobs indicate the photons that hit each DOM: the bigger, the higher the number of detected  Cherenkov photons. The color scale indicates the time from the first hit in the detector (light) to the last (dark).}
    \label{fig:NGCEvents}
\end{figure}
\begin{figure}
    \centering
    \includegraphics[width=0.4\linewidth]{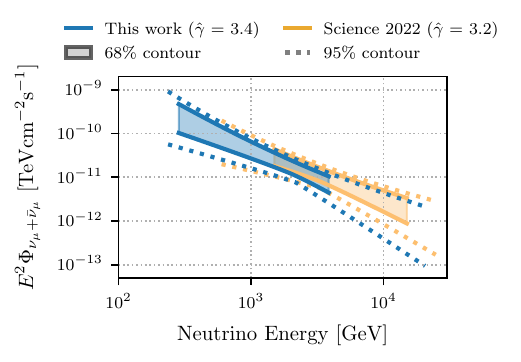}
    \caption{Comparison between the 68\% (solid) and 95\% (dashed) $\nu_\mu$ flux uncertainties and energy ranges of NGC\,1068's measurement of this work (blue) and \cite{IceCube:2022Science} (yellow). The 68\% energy interval obtained for this analysis ranges between 0.3 and 3.9\,TeV, while the 95\% one ranges between 0.2 and 20.6\,TeV, whilst the previous result spanned 1.5 and 15\,TeV for the 68\% contour and 0.6 and 24.8\,TeV for the 95\% one.
    }
    \label{fig:NGC_energy}
\end{figure}

In \autoref{sec:gamma-ray} we presented the result of a follow-up analysis on a list of 110 gamma-ray emitters, including NGC\,1068. The mean number of signal events $\hat{n}_\mathrm{s}$ contributing to the neutrino excess increased from 79 to 102, demonstrating that, despite the slight decrease in global significance of the analysis, new events contribute to the excess. In \autoref{fig:NGCEvents} we display 3 of the events contributing the most to the signal excess from NGC\,1068. On the left, we show the event contributing the most to the observed TS value obtained when testing NGC\,1068 for point-like neutrino emission.
In the middle, we show the most contributing event among the additional data that was included in this round of the analysis, which is the 9-th most contributing event. Finally, in the right panel, we display the most contributing event in IC79 data. This event is ranked 25-th among the most contributing to the excess.

In \autoref{fig:NGC_energy} we compare the newly measured spectrum of NGC\,1068 with that of \cite{IceCube:2022Science}. The energy ranges are computed by evaluating the TS contribution of each event adding to the excess, selecting events in the simulations with similar reconstructed energy and angular error, and building a histogram of their true neutrino energy. The histograms are then summed, weighted by their TS contribution. The displayed energy ranges correspond to the 68\% and 95\% central intervals of such a histogram. The flux intervals, instead, are computed assuming Wilks' theorem, which has been validated on Monte Carlo simulations. As already noted in \autoref{sec:gamma-ray}, this analysis reports a softer spectrum compared to the previous result (from $\hat\gamma = 3.2$ to $\hat\gamma = 3.4$) and a shift in the energy range towards lower energies. 
Despite the large overlap in the events contributing to NGC\,1068's neutrino excess in the two analyses, the new events are characterized by lower energies with respect to the ones driving the excess in \cite{IceCube:2022Science}, causing the observed change in energy range and best-fit spectral index. Nevertheless, the newly measured energy spectrum remains compatible with that of \cite{IceCube:2022Science} within the 95\% contours.

\begin{figure}
    \centering
    \includegraphics[width=0.3\linewidth]{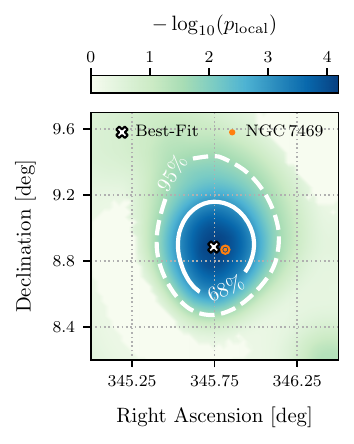}
    \caption{High-resolution scan and likelihood contours around the location of NGC\,7469, a type I Seyfert galaxy found to be the most significant source in the list of 47 X-ray bright AGN. Similarly to the right panel of \autoref{fig:skymap}, the white cross represents the best-fit location, the solid (dashed) line represents the 68\% (95\%) uncertainty contour of the excess, and the orange dot and circle represent, respectively, the source location and its optical size (obtained from the NASA/IPAC Extragalactic Database).}
    \label{fig:ngc7469}
\end{figure}

The search for neutrino emission from the list of 47 X-ray bright, non-blazar AGN, instead, found NGC\,7469 as the most significant, with 2.4\sigmas after accounting for the look-elsewhere effect. The likelihood fit for this source returned a very hard spectral index of $\hat\gamma = 1.9$ and a mean number of signal events of $\hat{n}_{\mathrm{s}} = 5.5$. As shown in \autoref{fig:ngc7469}, the likelihood scan in the vicinity of the source reveals that the best-fit position is well compatible with the cataloged coordinates. At the same time, the energy spectrum of this source, as opposed to NGC\,1068, spans energies of the order of 10 to 100 TeV (see \autoref{fig:spectra}). In fact, in this case, the neutrino excess is driven by only two very high-energy (E$_\nu > 100$ TeV) neutrino alert events (see also \citealt{Giacomo_NGC7569:ApJ2025}).

\begin{figure}
    \centering
    \includegraphics[width=0.8\linewidth]{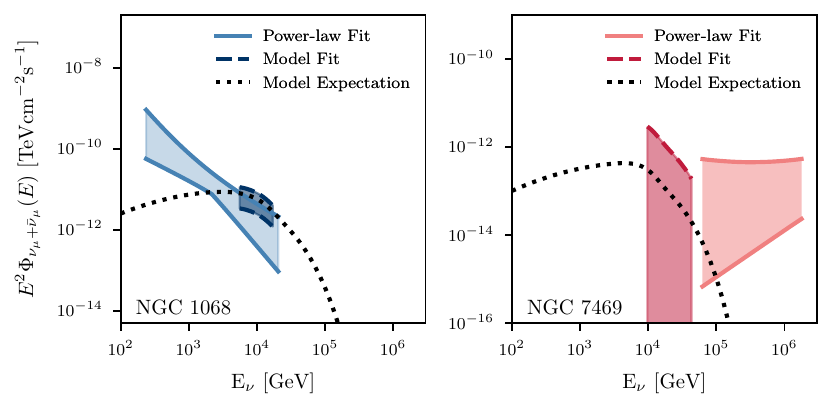}
    \caption{Comparison between the 95\% uncertainties and sensitive energy ranges of the power-law and core-corona model spectra for NGC\,1068 (left) and NGC\,7469 (right).  The power laws are shown in lighter colors with solid contours, while fluxes from the core-corona model are represented in darker colors with dashed contours. The dotted black lines represent the core-corona model expectations. The $\nu_{\mu}$ flux from NGC\,7469 in the core-corona model assumption is only partially constrained by the data.
    }
    \label{fig:fluxes_comparison}
\end{figure}

\begin{figure}
   \centering
   \includegraphics[width=0.4\linewidth]{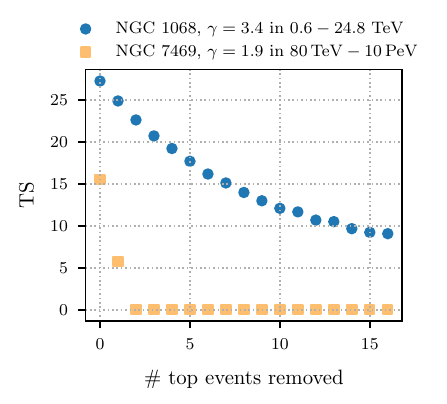}
   \caption{Dependency of the test statistics (TS) from the events with the highest $S/B$ ratio for the two most significant sources found in this analysis. We calculate the test statistics multiple times, removing the events with the highest $S/B$ one by one. While NGC\,1068's TS decreases slowly by removing the highest contributing events, NGC\,7469's TS rapidly drops to zero, as two high-energy events drive the excess.}
   \label{fig:TS_dependency}
\end{figure}

The emergence of a source with a markedly different spectrum than the well-known soft-spectrum source NGC\,1068 demonstrates the flexibility of the likelihood method. The custom core-corona model, which was also tested in this context, peaks at $\lesssim 10$\,TeV with a sharp cutoff at higher energies \citep{Icecube:2024Seyfert}. By construction, it penalizes any high-energy emission from the target source, as the right panel of \autoref{fig:fluxes_comparison} shows. \autoref{fig:TS_dependency} shows how the test statistic depends on the events with the highest signal-over-background ($S/B$) ratio. For NGC\,1068, removing the most significant events one by one leads to a gradual reduction in the test statistic, which eventually reaches zero. In contrast, for NGC\,7469, the test statistic drops to zero immediately after removing the two most significant events. This highlights the ability of the analysis to recover signals in both scenarios: a steady accumulation of many low-energy events from a soft-spectrum source like NGC\,1068, and a few high-energy events from a hard-spectrum source like NGC\,7469.
The starkly different behavior of these two sources also suggests that not all X-ray bright AGN share the same cosmic-ray acceleration and/or neutrino production mechanisms.

\subsection{Candidate source lists}\label{subsec:catalog_results}
Unlike the all-sky searches and the binomial tests, the catalog search on the list of 110 gamma-ray emitters was performed only under the assumption of a floating spectral index. This is the only case in which the global significance is computed analytically, using the well-known \v{S}id\'ak correction \citep{Sidak_correction}: $p_{\text{post}} = 1 - (1-p_{\text{local}})^N$, where $N$, in this case, is 110, i.e., the number of tested sources. No correlation between tests is included in the global significance calculation, since all sources are sufficiently spatially separated to be treated as independent.

The only catalog search not discussed in the main text is the one carried out on the list of 47 bright X-ray AGN, assuming neutrino emission follows the core-corona model \citep{Kheirandish:2021ApJ}. In this case, the global significance cannot be obtained with the \v{S}id\'ak correction alone, since we tested two spectral assumptions which might share some level of correlation. Therefore, after correcting for having tested 47 sources, we calculate the distribution of the minimum p-values out of pseudo-experiments analyzed under both spectral hypotheses and define the globally corrected $p$-value as the fraction of pseudo-experiments resulting in a $p$-value smaller than the source-corrected $p$-value. 

As in the case of the binomial test, the catalog search yielded a lower significance under the core-corona hypothesis compared to the power-law assumption. This outcome reinforces the conclusion that, given our current understanding of neutrino emission, the simple yet flexible assumption of an unbroken power-law spectrum remains the most consistent with the data.

In agreement with \cite{Icecube:2024Seyfert}, CGCG\,420-015 emerged as the most significant source in the list under the core-corona model, with a fitted signal strength of $\hat{n}_\mathrm{s} = 33.3$ and a local significance of 3.8\sigmas. NGC\,1068, the most prominent neutrino emitter in the sky, is always evaluated a posteriori. In this case, it was found to exhibit an excess of 48 signal events with a local significance of 5\sigmas.

A complete summary of all fit results is provided in \autoref{tab:gammaray} and \autoref{tab:xray}. These include the likelihood fit values, local significances, and 90\% upper limits for both $\gamma = 2$ and $\gamma = 3$ for all tested sources. In the second table, which presents results for the X-ray bright AGN sample, we additionally report the outcomes obtained under the core-corona model.

\begin{longtable*}{lrrrrrrr}
\caption{\textbf{List of gamma-ray emitters selected as candidate neutrino sources.} Sources
are ordered by descending significance. For each source, we list the name, the
equatorial coordinates (J2000 equinox) from the 4FGL-DR2 catalog \citep{4FGL:2020ApJS}, and likelihood search results: number of signal events $\hat{n}_{\mathrm{s}}$, spectral index $\hat\gamma$, $-\log_{10} p_{\mathrm{local}}$ and corresponding significance in brackets, 90\% C.L. astrophysical flux upper limits ($\phi_{90\%}$) with $\phi_{\nu_{\mu}+\bar{\nu}_{\mu}}=\phi_{90\%}(E_\nu/1\,\mathrm{TeV})^{-\gamma} \times 10^{-13}\,\mathrm{TeV^{-1} cm^{-2} s^{-1}}$ for $\gamma = 2.0$ and 3.0.}
\label{tab:gammaray} \\
\toprule
Source Name & R.A. & Dec. & $\hat{n}_\mathrm{s}$ & $\hat\gamma$ & $-\log_{10} p_{\mathrm{local}}$ & $\phi_{90\%}^{E^{-2}}$ & $\phi_{90\%}^{E^{-3}}$ \\
\midrule
\endfirsthead
\toprule
Source Name & R.A. & Dec. & $\hat{n}_\mathrm{s}$ & $\hat\gamma$ & $-\log_{10} p_{\mathrm{local}}$ & $\phi_{90\%}^{E^{-2}}$ & $\phi_{90\%}^{E^{-3}}$ \\
\midrule
\endhead
\midrule
\multicolumn{8}{r}{Continued on next page} \\
\midrule
\endfoot
\bottomrule
\endlastfoot

NGC\,1068 & 40.67 & $-$0.01 & 102 & 3.4 & $6.6\,(5.0\sigmas)$ & 6.6 & 385.6 \\
PKS\,1424+240 & 216.76 & 23.80 & 96 & 3.6 & $4.1\,(3.8\sigmas)$ & 8.3 & 227.2 \\
TXS\,0506+056 & 77.36 & 5.70 & 5 & 1.9 & $3.4\,(3.3\sigmas)$ & 5.2 & 249.5 \\
GB6\,J1542+6129 & 235.75 & 61.50 & 27 & 3.2 & $1.9\,(2.2\sigmas)$ & 9.3 & 145.8 \\
LQAC\,284+003 & 284.48 & 3.22 & 14 & 2.6 & $1.6\,(1.9\sigmas)$ & 3.4 & 178.5 \\
S3\,0458-02 & 75.30 & $-$1.97 & 34 & 4.4 & $1.4\,(1.8\sigmas)$ & 2.9 & 180.2 \\
OJ\,014 & 122.86 & 1.78 & 49 & 4.0 & $1.4\,(1.8\sigmas)$ & 3.0 & 173.9 \\
MITG\,J201534+3710 & 303.89 & 37.18 & 45 & 3.8 & $1.3\,(1.7\sigmas)$ & 5.6 & 119.4 \\
B2\,0619+33 & 95.73 & 33.43 & 48 & 4.4 & $1.3\,(1.6\sigmas)$ & 5.2 & 117.5 \\
MGRO\,J1908+06 & 286.91 & 6.32 & 2 & 1.8 & $1.3\,(1.6\sigmas)$ & 3.3 & 152.0 \\
NGC\,5380 & 209.33 & 37.50 & 12 & 2.7 & $1.2\,(1.6\sigmas)$ & 5.4 & 115.2 \\
B2\,1215+30 & 184.48 & 30.12 & 20 & 3.0 & $1.2\,(1.6\sigmas)$ & 4.9 & 117.6 \\
NGC\,2146 & 94.53 & 78.33 & 2 & 1.3 & $1.2\,(1.6\sigmas)$ & 9.7 & 138.6 \\
B3\,0609+413 & 93.22 & 41.37 & 9 & 2.4 & $1.2\,(1.5\sigmas)$ & 5.5 & 110.9 \\
S2\,0109+22 & 18.03 & 22.75 & 23 & 3.0 & $1.1\,(1.4\sigmas)$ & 4.1 & 117.3 \\
3C\,454.3 & 343.50 & 16.15 & 2 & 1.6 & $1.1\,(1.4\sigmas)$ & 3.7 & 124.9 \\
B2\,2234+28A & 339.10 & 28.48 & 28 & 3.3 & $1.0\,(1.3\sigmas)$ & 4.5 & 109.5 \\
TXS\,0603+476 & 91.86 & 47.66 & 35 & 4.4 & $1.0\,(1.3\sigmas)$ & 5.5 & 102.7 \\
S5\,1044+71 & 162.11 & 71.73 & 45 & 4.4 & $0.9\,(1.2\sigmas)$ & 8.6 & 111.1 \\
M\,31 & 10.82 & 41.24 & 20 & 3.5 & $0.9\,(1.2\sigmas)$ & 4.8 & 96.1 \\
PKS\,1441+25 & 220.99 & 25.03 & 6 & 2.2 & $0.9\,(1.2\sigmas)$ & 3.9 & 106.1 \\
3C\,273 & 187.27 & 2.05 & 34 & 4.4 & $0.9\,(1.1\sigmas)$ & 2.4 & 129.0 \\
7C\,2010+4619 & 303.02 & 46.49 & 4 & 2.0 & $0.8\,(1.1\sigmas)$ & 4.8 & 90.1 \\
PKS\,1502+106 & 226.10 & 10.50 & 7 & 2.3 & $0.8\,(1.0\sigmas)$ & 2.9 & 114.5 \\
B2\,2308+34 & 347.77 & 34.42 & 27 & 3.8 & $0.8\,(1.0\sigmas)$ & 4.2 & 90.3 \\
TXS\,0518+211 & 80.44 & 21.21 & 14 & 2.9 & $0.8\,(0.9\sigmas)$ & 3.3 & 93.3 \\
PKS\,1717+177 & 259.81 & 17.75 & 28 & 3.7 & $0.8\,(0.9\sigmas)$ & 3.2 & 99.2 \\
PMN\,J0948+0022 & 147.24 & 0.37 & 23 & 4.4 & $0.7\,(0.9\sigmas)$ & 2.1 & 120.0 \\
TXS\,1902+556 & 285.81 & 55.68 & 18 & 3.7 & $0.7\,(0.9\sigmas)$ & 4.7 & 80.6 \\
IC\,678 & 168.56 & 6.63 & 15 & 2.8 & $0.7\,(0.8\sigmas)$ & 2.5 & 106.6 \\
4C\,+55.17 & 149.42 & 55.38 & 18 & 3.4 & $0.7\,(0.8\sigmas)$ & 4.7 & 81.5 \\
4C\,+38.41 & 248.82 & 38.14 & 4 & 2.4 & $0.7\,(0.8\sigmas)$ & 3.9 & 79.3 \\
PKS\,0215+015 & 34.46 & 1.73 & 24 & 4.0 & $0.7\,(0.8\sigmas)$ & 2.1 & 111.7 \\
Mkn\,421 & 166.12 & 38.21 & 12 & 2.8 & $0.7\,(0.8\sigmas)$ & 3.9 & 78.9 \\
NVSS\,J141826-023336 & 214.61 & $-$2.56 & 11 & 4.4 & $0.6\,(0.8\sigmas)$ & 2.0 & 120.9 \\
OJ\,287 & 133.71 & 20.12 & 24 & 4.4 & $0.6\,(0.7\sigmas)$ & 3.0 & 84.5 \\
SBS\,0846+513 & 132.51 & 51.14 & 7 & 3.1 & $0.6\,(0.7\sigmas)$ & 4.2 & 72.1 \\
1ES\,0647+250 & 102.70 & 25.05 & 23 & 4.4 & $0.6\,(0.6\sigmas)$ & 3.1 & 81.7 \\
PKS\,B1130+008 & 173.20 & 0.57 & 17 & 4.4 & $0.6\,(0.6\sigmas)$ & 1.9 & 102.1 \\
1ES\,1959+650 & 300.01 & 65.15 & 6 & 2.8 & $0.6\,(0.6\sigmas)$ & 5.6 & 77.1 \\
BL\,Lac & 330.69 & 42.28 & 16 & 3.7 & $0.6\,(0.6\sigmas)$ & 3.7 & 72.1 \\
PKS\,0235+164 & 39.67 & 16.62 & 20 & 3.8 & $0.5\,(0.5\sigmas)$ & 2.6 & 79.6 \\
B2\,0218+357 & 35.28 & 35.94 & 14 & 4.4 & $0.4\,(0.4\sigmas)$ & 3.2 & 61.6 \\
MITG\,J200112+4352 & 300.30 & 43.89 & 7 & 3.0 & $0.4\,(0.4\sigmas)$ & 3.4 & 62.6 \\
Ton\,599 & 179.88 & 29.24 & 14 & 4.4 & $0.4\,(0.3\sigmas)$ & 2.9 & 65.0 \\
NGC\,1275 & 49.96 & 41.51 & 7 & 3.2 & $0.4\,(0.3\sigmas)$ & 3.2 & 59.7 \\
TXS\,1055+567 & 164.67 & 56.46 & 5 & 3.5 & $0.4\,(0.3\sigmas)$ & 3.7 & 57.3 \\
RX\,J1931.1+0937 & 292.78 & 9.63 & 12 & 4.4 & $0.4\,(0.3\sigmas)$ & 2.1 & 76.6 \\
4C\,+14.23 & 111.32 & 14.42 & 12 & 3.8 & $0.4\,(0.3\sigmas)$ & 2.2 & 69.5 \\
Arp\,299 & 172.07 & 58.52 & 8 & 4.4 & $0.4\,(0.3\sigmas)$ & 3.7 & 55.5 \\
Mkn\,501 & 253.47 & 39.76 & 9 & 4.4 & $0.4\,(0.2\sigmas)$ & 3.0 & 55.1 \\
S4\,1250+53 & 193.31 & 53.02 & 5 & 4.4 & $0.4\,(0.2\sigmas)$ & 3.4 & 51.2 \\
4C\,+15.54 & 241.77 & 15.84 & 10 & 4.4 & $0.4\,(0.2\sigmas)$ & 2.1 & 62.3 \\
NVSS\,J184425+154646 & 281.12 & 15.79 & 7 & 4.4 & $0.3\,(0.1\sigmas)$ & 2.1 & 61.0 \\
B2\,2114+33 & 319.06 & 33.66 & 3 & 2.8 & $0.3\,(0.1\sigmas)$ & 2.7 & 50.6 \\
4C\,+28.07 & 39.47 & 28.80 & 3 & 2.9 & $0.3\,(0.0\sigmas)$ & 2.5 & 51.1 \\
B3\,1343+451 & 206.39 & 44.88 & 2 & 3.0 & $0.3\,(0.0\sigmas)$ & 2.9 & 47.6 \\
4C\,+41.11 & 65.98 & 41.83 & 1 & 3.0 & $0.3\,(0.0\sigmas)$ & 2.8 & 45.5 \\
PKS\,0735+17 & 114.54 & 17.71 & 3 & 3.0 & $0.3\,(0.0\sigmas)$ & 2.1 & 54.7 \\
OX\,169 & 325.89 & 17.73 & 4 & 4.4 & $0.3\,(0.0\sigmas)$ & 2.0 & 53.1 \\
M\,82 & 148.95 & 69.67 & 4 & 4.4 & $0.3\,(0.0\sigmas)$ & 4.1 & 47.0 \\
MG2\,J043337+2905 & 68.41 & 29.10 & 1 & 4.4 & $0.3\,(0.0\sigmas)$ & 2.3 & 43.5 \\
W\,Comae & 185.38 & 28.24 & 0 & - & $0.3\,(0.0\sigmas)$ & 2.2 & 44.5 \\
S5\,1803+784 & 270.17 & 78.47 & 0 & - & $0.0\,(0.0\sigmas)$ & 6.1 & 80.9 \\
PMN\,J0709-0255 & 107.45 & $-$2.93 & 0 & - & $0.0\,(0.0\sigmas)$ & 1.6 & 80.9 \\
PKS\,1216-010 & 184.64 & $-$1.33 & 0 & - & $0.0\,(0.0\sigmas)$ & 1.4 & 66.7 \\
PKS\,0422+00 & 66.19 & 0.60 & 0 & - & $0.0\,(0.0\sigmas)$ & 1.4 & 67.1 \\
PKS\,0420-01 & 65.83 & $-$1.33 & 0 & - & $0.0\,(0.0\sigmas)$ & 1.4 & 68.1 \\
OT\,081 & 267.88 & 9.65 & 0 & - & $0.0\,(0.0\sigmas)$ & 1.8 & 58.6 \\
CTA\,102 & 338.15 & 11.73 & 0 & - & $0.0\,(0.0\sigmas)$ & 1.7 & 53.1 \\
PKS\,0336-01 & 54.88 & $-$1.78 & 0 & - & $0.0\,(0.0\sigmas)$ & 1.5 & 74.3 \\
PKS\,0736+01 & 114.82 & 1.62 & 0 & - & $0.0\,(0.0\sigmas)$ & 1.4 & 62.1 \\
PKS\,0440-00 & 70.66 & $-$0.30 & 0 & - & $0.0\,(0.0\sigmas)$ & 1.4 & 62.5 \\
1H\,1720+117 & 261.27 & 11.87 & 0 & - & $0.0\,(0.0\sigmas)$ & 1.7 & 52.2 \\
4C\,+01.28 & 164.62 & 1.56 & 0 & - & $0.0\,(0.0\sigmas)$ & 1.4 & 62.9 \\
M\,87 & 187.71 & 12.39 & 0 & - & $0.0\,(0.0\sigmas)$ & 1.8 & 54.4 \\
4C\,+01.02 & 17.17 & 1.58 & 0 & - & $0.0\,(0.0\sigmas)$ & 1.4 & 64.5 \\
OG\,050 & 83.17 & 7.55 & 0 & - & $0.0\,(0.0\sigmas)$ & 1.6 & 58.9 \\
TXS\,0141+268 & 26.15 & 27.09 & 0 & - & $0.0\,(0.0\sigmas)$ & 2.2 & 47.1 \\
RGB\,J2243+203 & 340.99 & 20.36 & 0 & - & $0.0\,(0.0\sigmas)$ & 2.0 & 48.3 \\
PKS\,0829+046 & 127.97 & 4.49 & 0 & - & $0.0\,(0.0\sigmas)$ & 1.6 & 61.3 \\
MITG\,J123931+0443 & 189.89 & 4.73 & 0 & - & $0.0\,(0.0\sigmas)$ & 1.5 & 59.2 \\
PG\,1553+113 & 238.93 & 11.19 & 0 & - & $0.0\,(0.0\sigmas)$ & 1.7 & 53.4 \\
PKS\,2032+107 & 308.85 & 10.94 & 0 & - & $0.0\,(0.0\sigmas)$ & 1.7 & 53.4 \\
4C\,+21.35 & 186.23 & 21.38 & 0 & - & $0.0\,(0.0\sigmas)$ & 1.9 & 47.0 \\
PG\,1218+304 & 185.34 & 30.17 & 0 & - & $0.0\,(0.0\sigmas)$ & 2.2 & 45.5 \\
RX\,J1754.1+3212 & 268.55 & 32.20 & 0 & - & $0.0\,(0.0\sigmas)$ & 2.3 & 45.8 \\
Arp\,220 & 233.70 & 23.53 & 0 & - & $0.0\,(0.0\sigmas)$ & 2.1 & 46.8 \\
MITG\,J021114+1051 & 32.81 & 10.86 & 0 & - & $0.0\,(0.0\sigmas)$ & 1.7 & 52.7 \\
87GB\,194024.3+102612 & 295.70 & 10.56 & 0 & - & $0.0\,(0.0\sigmas)$ & 1.7 & 54.4 \\
PKS\,0502+049 & 76.34 & 5.00 & 0 & - & $0.0\,(0.0\sigmas)$ & 1.5 & 56.7 \\
1ES\,0033+595 & 8.98 & 59.83 & 0 & - & $0.0\,(0.0\sigmas)$ & 3.6 & 49.2 \\
NGC\,3424 & 162.91 & 32.89 & 0 & - & $0.0\,(0.0\sigmas)$ & 2.3 & 45.5 \\
GB6\,J1037+5711 & 159.43 & 57.19 & 0 & - & $0.0\,(0.0\sigmas)$ & 3.3 & 46.8 \\
ON\,246 & 187.56 & 25.30 & 0 & - & $0.0\,(0.0\sigmas)$ & 2.1 & 43.4 \\
PKS\,1502+036 & 226.27 & 3.45 & 0 & - & $0.0\,(0.0\sigmas)$ & 1.5 & 59.0 \\
PKS\,0507+17 & 77.52 & 18.01 & 0 & - & $0.0\,(0.0\sigmas)$ & 1.8 & 47.0 \\
B2\,1520+31 & 230.55 & 31.74 & 0 & - & $0.0\,(0.0\sigmas)$ & 2.3 & 45.2 \\
PG\,1246+586 & 192.08 & 58.34 & 0 & - & $0.0\,(0.0\sigmas)$ & 3.3 & 46.3 \\
1ES\,0806+524 & 122.46 & 52.31 & 0 & - & $0.0\,(0.0\sigmas)$ & 3.1 & 45.8 \\
PKS\,0019+058 & 5.64 & 6.13 & 0 & - & $0.0\,(0.0\sigmas)$ & 1.5 & 56.0 \\
TXS\,2241+406 & 341.06 & 40.96 & 0 & - & $0.0\,(0.0\sigmas)$ & 2.6 & 44.6 \\
TXS\,1452+516 & 223.62 & 51.41 & 0 & - & $0.0\,(0.0\sigmas)$ & 3.0 & 43.7 \\
1H\,1013+498 & 153.77 & 49.43 & 0 & - & $0.0\,(0.0\sigmas)$ & 2.9 & 44.8 \\
B3\,0133+388 & 24.14 & 39.10 & 0 & - & $0.0\,(0.0\sigmas)$ & 2.5 & 44.0 \\
S4\,0814+42 & 124.56 & 42.38 & 0 & - & $0.0\,(0.0\sigmas)$ & 2.6 & 42.4 \\
S4\,0917+44 & 140.23 & 44.70 & 0 & - & $0.0\,(0.0\sigmas)$ & 2.6 & 40.7 \\
3C\,66A & 35.67 & 43.04 & 0 & - & $0.0\,(0.0\sigmas)$ & 2.6 & 43.0 \\
S4\,1749+70 & 267.16 & 70.10 & 0 & - & $0.0\,(0.0\sigmas)$ & 3.7 & 41.6 \\
S5\,0716+71 & 110.49 & 71.34 & 0 & - & $0.0\,(0.0\sigmas)$ & 3.5 & 37.9 \\
\end{longtable*}

\bigskip
\bigskip

\begin{turnpage}
\centering
\begin{longtable*}{lrrr|rrrrr|rrr}
\caption{\textbf{List of X-ray-bright AGN selected as candidate neutrino sources.} Sources
are ordered by descending significance under the power-law spectral assumption. For each source we list the name, the equatorial coordinates (J2000 equinox) from the BASS catalog \citep{Koss:2022ApJS}, intrinsic 20--50~keV X-ray flux ($F^{\mathrm{intr}}_{20\text{--}50\,\mathrm{keV}}\,\times 10^{-11}\,\mathrm{erg\,cm^{-2}s^{-1}}$), the likelihood search results under the power-law hypothesis: $\hat{n}_{\mathrm{s}}$, spectral index $\hat\gamma$, $-\log_{10} p_{\mathrm{local}}$ and corresponding significance in brackets, 90\% C.L. astrophysical flux upper limits ($\phi_{90\%}$) with $\phi_{\nu_{\mu}+\bar{\nu}_{\mu}}=\phi_{90\%}(E_\nu/1\,\mathrm{TeV})^{-\gamma} \times 10^{-13}\,\mathrm{TeV^{-1} cm^{-2} s^{-1}}$ for $\gamma = 2.0$ and 3.0., and the results for the core-corona model hypothesis: number of signal events $\hat{n}_{\mathrm{s}}$, $-\log_{10} p_{\mathrm{local}}$ and corresponding significance in brackets, and 90\% C.L. astrophysical flux upper limits given as mean number of signal events. The sources highlighted with an $^*$ were included in the \citet{Icecube:2024Seyfert} selection.}
\label{tab:xray} \\
\toprule
\multicolumn{4}{l|}{} & \multicolumn{5}{c}{Power-law Fit Results} & \multicolumn{3}{|c}{Core-Corona Model Fit Results} \\
\midrule
Source Name & R.A. & Dec. & $F^{\mathrm{intr}}_{20-50\,\mathrm{keV}}$ & $\hat{n}_\mathrm{s}$ & $\hat\gamma$ & $-\log_{10} p_{\mathrm{local}}$ & $\phi_{90\%}^{E^{-2}}$ & $\phi_{90\%}^{E^{-3}}$ & $\hat{n}_\mathrm{s}$ & $-\log_{10} p_{\mathrm{local}}$ & ${n_\mathrm{s}}_{90\%}^{\text{Model}}$ \\
\midrule
\endfirsthead
\toprule
\multicolumn{4}{l|}{} & \multicolumn{5}{c}{Power-law Fit Results} & \multicolumn{3}{|c}{Core-Corona Model Fit Results} \\
\midrule
Source Name & R.A. & Dec. & $F^{\mathrm{intr}}_{20-50\,\mathrm{keV}}$ & $\hat{n}_\mathrm{s}$ & $\hat\gamma$ & $-\log_{10} p_{\mathrm{local}}$ & $\phi_{90\%}^{E^{-2}}$ & $\phi_{90\%}^{E^{-3}}$ & $\hat{n}_\mathrm{s}$ & $-\log_{10} p_{\mathrm{local}}$ & ${n_\mathrm{s}}_{90\%}^{\text{Model}}$ \\
\midrule
\endhead
\midrule
\multicolumn{12}{r}{Continued on next page} \\
\midrule
\endfoot
\bottomrule
\endlastfoot

NGC\,1068$^*$ & 40.67 & $-$0.01 & 7.72 & 102 & 3.4 & $6.6\,(5.0\sigmas)$ & 6.6 & 385.6 & 48 & $5.8\,(4.7\sigmas)$ & 73.0 \\
NGC\,7469 & 345.82 & 8.87 & 2.69 & 5 & 1.9 & $4.1\,(3.8\sigmas)$ & 6.3 & 265.7 & 7 & $0.9\,(1.1\sigmas)$ & 58.0 \\
NGC\,4151$^*$ & 182.64 & 39.41 & 18.09 & 28 & 2.7 & $2.9\,(3.1\sigmas)$ & 8.5 & 178.1 & 23 & $3.0\,(3.1\sigmas)$ & 45.5 \\
CGCG\,420-015$^*$ & 73.36 & 4.06 & 1.77 & 35 & 2.7 & $2.4\,(2.7\sigmas)$ & 4.3 & 222.1 & 33 & $3.4\,(3.4\sigmas)$ & 46.3 \\
Cygnus\,A$^*$ & 299.87 & 40.73 & 4.93 & 3 & 1.6 & $2.2\,(2.5\sigmas)$ & 7.5 & 153.8 & 1 & $0.5\,(0.5\sigmas)$ & 34.2 \\
LEDA\,166445 & 42.68 & 54.70 & 1.61 & 57 & 4.4 & $1.8\,(2.1\sigmas)$ & 7.6 & 136.9 & 0 & $0.0\,(0.0\sigmas)$ & 30.4 \\
NGC\,4992 & 197.27 & 11.63 & 2.34 & 27 & 2.9 & $1.6\,(2.0\sigmas)$ & 4.1 & 162.0 & 22 & $2.5\,(2.7\sigmas)$ & 35.7 \\
NGC\,1194$^*$ & 45.95 & $-$1.10 & 3.87 & 43 & 4.4 & $1.5\,(1.8\sigmas)$ & 3.0 & 182.0 & 1 & $0.4\,(0.3\sigmas)$ & 32.8 \\
Mrk\,1498 & 247.02 & 51.78 & 1.86 & 40 & 3.6 & $1.4\,(1.7\sigmas)$ & 6.5 & 120.5 & 8 & $1.0\,(1.3\sigmas)$ & 25.3 \\
MCG\,+4-48-2$^*$ & 307.15 & 25.73 & 4.32 & 37 & 3.2 & $1.4\,(1.7\sigmas)$ & 4.9 & 130.3 & 17 & $1.6\,(2.0\sigmas)$ & 31.0 \\
NGC\,3079 & 150.49 & 55.68 & 3.33 & 34 & 3.6 & $1.3\,(1.7\sigmas)$ & 6.6 & 117.2 & 16 & $1.3\,(1.6\sigmas)$ & 36.9 \\
Mrk\,417 & 162.38 & 22.96 & 1.73 & 4 & 1.9 & $1.3\,(1.6\sigmas)$ & 4.4 & 127.6 & 5 & $1.0\,(1.3\sigmas)$ & 28.3 \\
Q0241+622 & 41.24 & 62.47 & 3.45 & 17 & 2.8 & $1.2\,(1.6\sigmas)$ & 7.8 & 118.7 & 14 & $1.9\,(2.3\sigmas)$ & 23.6 \\
LEDA\,138501 & 32.41 & 52.44 & 1.95 & 34 & 4.4 & $1.0\,(1.2\sigmas)$ & 5.4 & 97.5 & 0 & $0.0\,(0.0\sigmas)$ & 21.2 \\
LEDA\,86269 & 71.04 & 28.22 & 1.76 & 39 & 4.4 & $1.0\,(1.2\sigmas)$ & 4.3 & 104.5 & 0 & $0.0\,(0.0\sigmas)$ & 26.3 \\
NGC\,5252 & 204.57 & 4.54 & 3.65 & 32 & 3.6 & $0.9\,(1.1\sigmas)$ & 2.6 & 125.4 & 7 & $0.7\,(0.9\sigmas)$ & 28.1 \\
3C\,382$^*$ & 278.76 & 32.70 & 2.62 & 33 & 4.4 & $0.9\,(1.1\sigmas)$ & 4.2 & 93.7 & 0 & $0.0\,(0.0\sigmas)$ & 23.3 \\
NGC\,4388$^*$ & 186.44 & 12.66 & 10.79 & 2 & 2.0 & $0.8\,(1.0\sigmas)$ & 3.0 & 111.2 & 0 & $0.0\,(0.0\sigmas)$ & 26.0 \\
LEDA\,168563 & 73.02 & 49.55 & 2.15 & 2 & 1.8 & $0.8\,(1.0\sigmas)$ & 4.9 & 88.3 & 2 & $0.6\,(0.7\sigmas)$ & 19.5 \\
Ark\,120 & 79.05 & $-$0.15 & 2.75 & 27 & 4.4 & $0.8\,(0.9\sigmas)$ & 2.2 & 124.4 & 0 & $0.0\,(0.0\sigmas)$ & 24.3 \\
Z164-19$^*$ & 221.40 & 27.03 & 8.81 & 3 & 1.9 & $0.7\,(0.9\sigmas)$ & 3.6 & 89.5 & 0 & $0.0\,(0.0\sigmas)$ & 22.5 \\
3C\,445 & 335.96 & $-$2.10 & 2.02 & 14 & 4.4 & $0.7\,(0.8\sigmas)$ & 2.0 & 124.1 & 2 & $0.5\,(0.5\sigmas)$ & 20.4 \\
NGC\,5548 & 214.50 & 25.14 & 2.70 & 17 & 3.2 & $0.7\,(0.8\sigmas)$ & 3.4 & 85.8 & 7 & $0.9\,(1.2\sigmas)$ & 22.1 \\
Mrk\,6 & 103.05 & 74.43 & 2.15 & 10 & 2.8 & $0.6\,(0.7\sigmas)$ & 6.8 & 88.2 & 9 & $1.1\,(1.4\sigmas)$ & 19.1 \\
NGC\,3516$^*$ & 166.70 & 72.57 & 4.17 & 34 & 4.4 & $0.6\,(0.6\sigmas)$ & 6.8 & 86.7 & 0 & $0.0\,(0.0\sigmas)$ & 20.6 \\
UGC\,11910$^*$ & 331.76 & 10.23 & 2.20 & 21 & 4.4 & $0.6\,(0.6\sigmas)$ & 2.4 & 87.9 & 6 & $0.7\,(0.9\sigmas)$ & 21.6 \\
4C\,+50.55$^*$ & 321.16 & 50.97 & 7.73 & 9 & 3.0 & $0.6\,(0.6\sigmas)$ & 4.1 & 67.8 & 8 & $1.0\,(1.2\sigmas)$ & 16.2 \\
IGR\,J21277+5656 & 321.94 & 56.94 & 1.67 & 4 & 2.4 & $0.6\,(0.6\sigmas)$ & 4.3 & 67.9 & 6 & $0.9\,(1.1\sigmas)$ & 16.7 \\
Mrk\,1040$^*$ & 37.06 & 31.31 & 2.37 & 16 & 4.4 & $0.5\,(0.5\sigmas)$ & 3.1 & 69.7 & 3 & $0.6\,(0.7\sigmas)$ & 17.3 \\
3C\,111$^*$ & 64.59 & 38.03 & 4.13 & 12 & 4.4 & $0.4\,(0.3\sigmas)$ & 3.1 & 59.3 & 0 & $0.0\,(0.0\sigmas)$ & 15.3 \\
Mrk\,1210$^*$ & 121.02 & 5.11 & 2.36 & 9 & 4.4 & $0.4\,(0.1\sigmas)$ & 1.7 & 72.5 & 0 & $0.0\,(0.0\sigmas)$ & 18.4 \\
NGC\,1142 & 43.80 & $-$0.18 & 4.05 & 6 & 3.7 & $0.3\,(0.1\sigmas)$ & 1.5 & 75.7 & 2 & $0.4\,(0.3\sigmas)$ & 16.0 \\
NGC\,7682$^*$ & 352.27 & 3.53 & 1.99 & 1 & 2.8 & $0.3\,(0.1\sigmas)$ & 1.6 & 69.6 & 2 & $0.6\,(0.7\sigmas)$ & 16.9 \\
NGC\,7603 & 349.74 & 0.24 & 1.87 & 2 & 4.4 & $0.3\,(0.0\sigmas)$ & 1.5 & 68.4 & 0 & $0.0\,(0.0\sigmas)$ & 15.1 \\
3C\,390.3$^*$ & 280.54 & 79.77 & 3.66 & 0 & - & $0.0\,(0.0\sigmas)$ & 6.7 & 93.7 & 0 & $0.0\,(0.0\sigmas)$ & 13.4 \\
4C\,+74.26 & 310.66 & 75.13 & 2.00 & 0 & - & $0.0\,(0.0\sigmas)$ & 4.5 & 49.8 & 0 & $0.0\,(0.0\sigmas)$ & 12.1 \\
NGC\,6240$^*$ & 253.25 & 2.40 & 12.89 & 0 & - & $0.0\,(0.0\sigmas)$ & 1.4 & 57.4 & 0 & $0.0\,(0.0\sigmas)$ & 14.8 \\
NGC\,3227$^*$ & 155.88 & 19.87 & 4.16 & 0 & - & $0.0\,(0.0\sigmas)$ & 1.9 & 47.7 & 0 & $0.0\,(0.0\sigmas)$ & 15.4 \\
IRAS\,05589+2828 & 90.54 & 28.47 & 2.64 & 0 & - & $0.0\,(0.0\sigmas)$ & 2.2 & 42.8 & 0 & $0.0\,(0.0\sigmas)$ & 12.6 \\
Mrk\,79 & 115.64 & 49.81 & 1.82 & 0 & - & $0.0\,(0.0\sigmas)$ & 3.0 & 44.8 & 0 & $0.0\,(0.0\sigmas)$ & 12.1 \\
2MASX\,J20145928\\\ \ \ \ \ \ \ \ \ \ ~+2523010$^*$ & 303.75 & 25.38 & 2.70 & 0 & - & $0.0\,(0.0\sigmas)$ & 2.0 & 42.8 & 0 & $0.0\,(0.0\sigmas)$ & 13.1 \\
IRAS\,05078+1626$^*$ & 77.69 & 16.50 & 3.37 & 0 & - & $0.0\,(0.0\sigmas)$ & 1.8 & 46.0 & 0 & $0.0\,(0.0\sigmas)$ & 13.2 \\
Mrk\,110$^*$ & 141.30 & 52.29 & 2.12 & 0 & - & $0.0\,(0.0\sigmas)$ & 3.0 & 44.6 & 0 & $0.0\,(0.0\sigmas)$ & 11.7 \\
NGC\,4102 & 181.60 & 52.71 & 2.24 & 0 & - & $0.0\,(0.0\sigmas)$ & 3.1 & 46.5 & 0 & $0.0\,(0.0\sigmas)$ & 15.2 \\
NGC\,7319 & 339.01 & 33.98 & 1.70 & 0 & - & $0.0\,(0.0\sigmas)$ & 2.3 & 41.2 & 1 & $0.5\,(0.4\sigmas)$ & 12.7 \\
NGC\,4051 & 180.79 & 44.53 & 1.75 & 0 & - & $0.0\,(0.0\sigmas)$ & 2.7 & 43.1 & 0 & $0.0\,(0.0\sigmas)$ & 16.6 \\
UGC\,3374$^*$ & 88.72 & 46.44 & 4.94 & 0 & - & $0.0\,(0.0\sigmas)$ & 2.7 & 42.0 & 0 & $0.0\,(0.0\sigmas)$ & 11.7 \\
Mrk\,3$^*$ & 93.90 & 71.04 & 8.98 & 0 & - & $0.0\,(0.0\sigmas)$ & 3.5 & 39.7 & 0 & $0.0\,(0.0\sigmas)$ & 11.3 \\
\end{longtable*}

\end{turnpage}

\end{appendix}

\end{document}